\RequirePackage{fix-cm}
\documentclass[numbook]{svjour3}                     
\counterwithin*{section}{part}
\smartqed  

\usepackage{graphicx}
\usepackage{titlesec}
\usepackage{longtable}
\usepackage{algorithm}
\usepackage{algpseudocode}
\usepackage{tabularx}
\usepackage[comma,authoryear]{natbib}

\usepackage{array}
\newcolumntype{L}{>{\arraybackslash}m{12cm}}
\newcolumntype{P}{>{\arraybackslash}m{6cm}}
\newcolumntype{C}{>{\centering\arraybackslash}m{12cm}}

\usepackage{amsmath,amssymb,amsfonts} 
\usepackage{cite}
\usepackage{url}
\usepackage{relsize}
\usepackage{lscape}
\usepackage{multirow}
\usepackage{ragged2e}
\usepackage{adjustbox, tabularx,booktabs}
\usepackage{dcolumn}
\usepackage{float}
\floatstyle{plaintop}
\usepackage{adjustbox}
\restylefloat{table}
\usepackage{geometry}
\usepackage{graphicx}
\usepackage{algorithm}
\usepackage{algpseudocode}

\usepackage{booktabs} 
\usepackage{array} 
\usepackage{paralist} 
\usepackage{verbatim} 

\usepackage{ifthen}
\usepackage{xspace}
\usepackage[table]{xcolor}
\usepackage{fancybox} 
\usepackage{soul}
\usepackage{hyperref}
\usepackage{cleveref}
\usepackage{color, colortbl}
\usepackage{balance}

\algnewcommand\algorithmicforeach{\textbf{for each}}
\algdef{S}[FOR]{ForEach}[1]{\algorithmicforeach\ #1\ \algorithmicdo}

\newcommand\clearrow{\global\let\rowmac\relax}

\newboolean{includecomments}
\setboolean{includecomments}{true}

\newcommand{\nbq}[3]{%
	\colorbox{#3}{\bfseries\sffamily\scriptsize\textcolor{white}{#1}}
	\textcolor{#3}{\sf$\blacktriangleright$ \textit{#2} $\blacktriangleleft$}
}

\definecolor{colorTODO}{HTML}{0000FF}
\definecolor{colorGulsher}{HTML}{C90800}
\definecolor{colorAjibode}{HTML}{006400}

\ifthenelse{\boolean{includecomments}} {
	\newcommand{\TODO}[1]{\nbq{TODO}{#1}{colorTODO}}
	\newcommand{\Ajibode}[1]{\nbq{Ajibode}{#1}{colorAjibode}}
	\newcommand{\Gulsher}[1]{\nbq{Gulsher}{#1}{colorGulsher}}
}{
	\newcommand{\TODO}[1]{}
	\newcommand{\Ajibode}[1]{}
	\newcommand{\Gulsher}[1]{}
}

\def\RQOne{What are the most common issues reported in \ml libraries and how do they affect library maintenance?\xspace}

\def\RQTwo{Which related issue is mostly faced by the \ml library users?\xspace}

\def\RQThree{How does the number of comments on an issue affect the time it takes to resolve the issue in machine learning libraries?\xspace}

\def\RQFour{What is the average response time for issues reported in machine learning libraries?\xspace}

\def\RQFive{Are there any impacts that issues priority have on the duration of the issue?\xspace}

\def\RQSix{What are the status of the closed issues of each machine learning library?\xspace}

\def\RQSeven{How does the number of assignee/contributors of an issue affect the time it takes to close the issue in machine learning libraries?\xspace}

\def\etal{\textit{et al.}\xspace}

\def\git{GitHub\xspace}
\def\te{Tensorflow\xspace}
\def\ke{Keras\xspace}
\def\th{Theano\xspace}
\def\sk{Scikit-learn\xspace}
\def\py{Pytorch\xspace}
\def\ca{Caffe\xspace}
\def\ml{machine-learning\xspace}

\begin{document}

\title{Software issues report for bug fixing process: An empirical study of \ml libraries.}



\author{Ajibode Adekunle         \and
        Yunwei Dong $^*$ \thanks{$*$ Corresponding author} 
        \and Hongji Yang}


\institute{Ajibode Adekunle Akinjobi \at
              School of Software, Northwestern Polytechnical University, Xi'an, 710072, China\\
              \email{damajibode@gmail.com}
           \and
           Yunwei Dong \at
              School of Computer Science, Northwestern Polytechnical University, Xi'an 710072, China\\
              \email{yunweidong@nwpu.edu.cn}
           \and
           Hongji Yang \at
               School of Computing and Mathematical Sciences, Leicester University, Leicester, UK\\
              \email{hongji.yang@leicester.ac.uk}
}

\date{Received: date / Accepted: date}

\maketitle

\begin{abstract}
Issue resolution and bug-fixing processes are essential in the development of \ml libraries, similar to software development, to ensure well-optimized functions. Understanding the issue resolution and bug-fixing process of \ml libraries can help developers identify areas for improvement and optimize their strategies for issue resolution and bug-fixing. However, detailed studies on this topic are lacking. Therefore, we investigated the effectiveness of issue resolution for bug-fixing processes in six \ml libraries: \te, \ke, \th, \py, \ca, and \sk. We addressed seven research questions (RQs) using 16,921 issues extracted from the \git repository via the \git Rest API. We employed several quantitative methods of data analysis, including correlation, OLS regression, percentage and frequency count, and heatmap to analyze the RQs. We found the following through our empirical investigation: (1) The most common categories of issues that arise in \ml libraries are bugs, documentation, optimization, crashes, enhancement, new feature requests, build/CI, support, and performance. (2) Effective strategies for addressing these problems include fixing critical bugs, optimizing performance, and improving documentation. (3) These categorized issues are related to testing and runtime and are common among all six \ml libraries. (4) Monitoring the total number of comments on issues can provide insights into the duration of the issues. (5) It is crucial to strike a balance between prioritizing critical issues and addressing other issues in a timely manner. Therefore, this study concludes that efficient issue-tracking processes, effective communication, and collaboration are vital for effective resolution of issues and bug fixing processes in \ml libraries.

\end{abstract}

\section{Introduction}
\label{Sec::Introduction}
Software development and maintenance are critical activities of the modern software industry. Software systems must be developed, maintained, and updated to meet the evolution requirements of users and businesses goals. However, software development and maintenance are not without challenges, one of which is the issue of software bugs \citep{saha2015understanding}.

Similarly, \ml libraries are also developed and maintained like other software, and have become increasingly popular in recent years as a way to simplify the development of \ml applications. These libraries provide pre-built algorithms and functions that can be easily integrated into software applications. However, like any software system, \ml libraries are not immune to bugs and issues \citep{nguyen2019machine}.

To address the issue of bugs and general issues in \ml libraries, it is essential to have an effective issues resolution and bug fixing process \citep{thung2012empirical}. An issues report is an important component of the bug fixing process \citep{guo2011not, goyal2019empirical, zhang2015survey}. It provides a comprehensive overview of the issues and bugs in the software system, enabling developers to prioritize and fix them. 

These issues are reported in different repositories, such as \git, in a form of tickets or issues, which typically include a description of the problem, steps to reproduce the issue, and any relevant code or data. Users can also add comments and discuss possible solutions or workarounds. 

\git is an online platform used for collaborative software development and version control, currently holds the distinction of being the largest host of open-source software in the world. In the year 2018, the platform saw the creation of approximately one third of its total 96 million repositories, indicating the platform's continued growth and significance as a resource for open-source software development \citep{brisson2020we}.

Bug fixing process refers to the steps taken to identify and resolve software defects, with the goal of improving the quality and reliability of the software. This process involve activities such as bug detection, analysis, isolation, and correction, and may be influenced by factors such as the software development methodology, the skills and resources available, and the needs and expectations of users. Many research have been conducted in the area of bug fixing process \citep{sepahvand2022effective, sepahvand2020predicting, ozdaugouglu2019monitoring, goyal2020performance, francalanci2008empirical}. Majority of the studies focus more on analyzing the effectiveness of different bug fixing strategies and techniques \citep{liu2013r2fix, zhong2015empirical}, as well as identifying common challenges and issues faced by developers in the bug fixing process \citep{latoza2006maintaining, mezouar2018tweets}. Some studies have also explored the role of automated tools and techniques \citep{nguyen2010recurring}, such as \ml and natural language processing, in improving the efficiency and effectiveness of bug fixing.

However, there is no study that have revealed the issues resolution and bug fixing process of \ml libraries in detail. This is an important area of research, as \ml libraries play a critical role in the development of \ml applications, and any bugs or issues in these libraries can have a significant impact on the performance and reliability of these applications.

Understanding the issues resolution and bug fixing process of \ml libraries can help developers to identify areas for improvement and optimize their issues resolution and bug fixing strategies. This can also lead to the development of better tools and techniques for bug detection and resolution in \ml libraries.

Furthermore, software development and maintenance, including the development and maintenance of \ml libraries, are essential components of the modern software industry \citep{chen2022security, gartziandia2022machine}. To address the challenges of bugs and issues in these systems, an effective bug fixing process is necessary, which includes the reporting of issues through platforms such as \git. Similarly, research is needed to understand the issues status, the resolution speed, and bug fixing process of \ml libraries in detail and to develop more effective bug fixing strategies and users satisfaction for these critical systems.

Therefore, this study aims to conduct an empirical investigation of issues resolution for effective bug fixing process of \ml libraries through their various issues reports. We examine the issues reports generated for several popular \ml libraries, \te, \py, \ke, \ca, \sk, and \th, and analyze them to determine the effectiveness of their bug fixing process. This study contributes to a better understanding of the bug fixing process in \ml libraries, which can help developers and users of these libraries to identify potential issues and improve the overall quality of the software. In this study, we answer the following research questions (RQs):

\begin{itemize}
\item[\textbf{$RQ_1.$}] \textit{\RQOne}

The aim of this research question is to identify the most common issues reported in \ml libraries and understand how they affect library maintainers. Knowing the most common issues reported in \ml libraries is important for library maintainers because it helps them prioritize their bug fixing efforts. By understanding which issues are most likely to occur and which ones have the biggest impact on their users, maintainers can allocate their limited resources more effectively. Additionally, understanding how these issues affect maintainers (e.g., by increasing their workload or requiring specialized expertise) can help organizations plan for and mitigate these effects.

\item[\textbf{$RQ_2.$}] \textit{\RQTwo}

This question identifies the issues that \ml library users face most frequently. Understanding the related issues faced by \ml library users is important for library maintainers because it helps them to design better libraries. By understanding which related issues are most frustrating or impactful for users, maintainers can prioritize these issues in their bug fixing efforts, and design new features and improvements to address them. This RQ is different from RQ1 because we will understand what type of problems, such as \textit{runtime} problem, \textit{algorithm} problem, or any type of problem that the outcome of RQ1 related to. This research question also tells us which stage users encounter issues identified in RQ1. For example, if the outcome of RQ1 is "bug,” and the outcome of RQ2 is \textit{runtime}, we can conclude that the users encountered a bug during the \textit{runtime}. This can lead to better user experience and increased adoption of the library.

\item[\textbf{$RQ_3.$}] \textit{\RQThree}

This question helps to understand the relationship between the number of comments on an issue and the time it takes to resolve it. Understanding the relationship between the number of comments on an issue and the time it takes to resolve it is important for library maintainers because it helps them manage their bug tracking systems more effectively. By understanding the factors that influence issue resolution time, maintainers can optimize their workflows and identify bottlenecks in the process. This can help them reduce the time it takes to fix issues and improve overall developer productivity.

\item[\textbf{$RQ_4.$}] \textit{\RQFour}

This research question aims to determine the average response time for issues reported in \ml libraries. Knowing the average response time for issues reported in \ml libraries is important for both library maintainers and users. For maintainers, it provides a metric for measuring their responsiveness to user needs and can help them identify areas for improvement. For users, it can help set expectations for when they can expect their issues to be addressed and provide feedback on the quality of support they receive.

\item[\textbf{$RQ_5.$}] \textit{\RQFive}

The aim of this research question is to understand how issue priority affects the duration of issue resolution. Understanding the impact of issue priority on the duration of issue resolution is important for library maintainers because it helps them prioritize their bug fixing efforts. By understanding how issue priority affects resolution time, maintainers can ensure that the most critical issues are addressed first, which can help mitigate the impact of bugs on users and improve the overall quality of the library.

\item[\textbf{$RQ_6.$}] \textit{\RQSix}

This research question is set out to understand the states of closed issues in \ml libraries. Knowing the states of closed issues in \ml libraries is important for both library maintainers and users. For maintainers, it provides insight into how well the bug fixing process is working and can help identify areas for improvement. For users, it can help set expectations for how likely it is that their issues will be resolved and can provide feedback on the quality of support they receive.

\item[\textbf{$RQ_7.$}] \textit{\RQSeven}

We aim to answer this research question to understand the relationship between the number of assignees/contributors on an issue and the time it takes to close it. Understanding the relationship between the number of assignees/contributors on an issue and the time it takes to close it is important for library maintainers because it helps them manage their bug fixing resources more effectively. By understanding the factors that influence issue resolution time, maintainers can optimize their workflows and identify opportunities to involve more developers in resolving issues. This can help reduce the time it takes to fix issues and improve overall developer productivity.
\end{itemize}

In order to address our research questions, we conducted an analysis on 16,921 issues that were reported in six different \ml libraries. The findings of the study revealed several noteworthy observations. For instance, a significant portion of the issues reported in the \ml libraries under examination were not properly categorized. However, the most frequently categorized issue in the studied libraries was the \textit{Bug} category, which appeared in five \ml libraries. This suggests that developers must concentrate on enhancing the quality of their libraries.

Moreover, our study discovered that the number of comments on an issue has a considerable impact on the duration of issue resolution. Unfortunately, prioritized issues received a higher number of comments as compared to unprioritized issues. Additionally, our research indicates that the majority of the \ml libraries analyzed exhibited a weak correlation between the number of contributors and the duration of issues, suggesting a minimal association between contributors and issues duration.

Drawing upon our empirical findings, we offer several crucial recommendations and implications for developers and library maintainers to improve the bug fixing process and issue resolution in their \ml libraries.
 
To summarize, this paper makes the following contributions:
\begin{itemize}
    \item This paper presents an empirical study on issues reported in \ml libraries, which offers several important contributions to the field. Firstly, we provide the first comprehensive analysis of issues across six major \ml libraries, spanning a total of 16,921 reported issues within the last five years. This dataset represents a valuable resource for understanding the current state of \ml library maintenance and can be used as a baseline for future research.
    \item Our study reveals that \ml libraries are facing significant challenges in terms of issue resolution and maintenance, highlighting the need for better tools and strategies for issue classification, prioritization, and status updates. By identifying key issues and providing recommendations for improving the management and resolution of software issues, we help library maintainers and software developers alike improve the quality and reliability of \ml libraries.
    \item Our study makes an important contribution to open science by publicly sharing the datasets used in our analysis. This is available in our replication package \footnote{https://bitbucket.org/eyinlojuoluwa/issue-resolution/downloads/}. This will enable other researchers to replicate our findings, extend our analysis to other libraries or domains, and contribute to the ongoing efforts to improve the quality and reliability of software systems. Overall, this study represents a significant step forward in our understanding of issues in \ml libraries and provides a roadmap for future research and development in this area.
\end{itemize}

This paper is organized as follows: \Cref{Sec::motivation} outlines the motivation for the study, and \Cref{Sec::concepts} details the relevant concepts and terminologies pertaining to issue resolution and the bug-fixing process. \Cref{Sec::methodology} describes the methodology employed for data collection and analysis to address our research questions. \Cref{Sec::result} presents the results obtained. We further discuss the findings of this study in \Cref{Sec::discussion} and the limitations and potential threats to the validity of our study in \Cref{Sec::ttv}. A review of the relevant prior research is presented in \Cref{Sec::related}. Finally, we draw our conclusions in \Cref{Sec::conclusion}.


\section{Motivation}
\label{Sec::motivation}
Software development has become an essential part of our daily lives, and as such, software engineers continually strive to improve the quality of software products \citep{gheorghe2020agile}. Despite the best efforts of developers, software bugs are common during the software development process. Defects, also known as bugs, can cause significant damage to software systems, leading to system crashes, data losses, and other errors \citep{kaur2022empirical}. To mitigate the impact of defects, it is essential to identify, classify, and resolve them as quickly as possible \citep{fradrich2020common}.

One way to manage software bugs is through the use of issue reporting systems \citep{yu2016reviewer, yu2014empirical, liao2018exploring}. An issue-reporting system is a platform where software users and developers can report and track software issues. Software developers use these reports to diagnose and resolve bugs, thereby ensuring that the software remains functional and reliable.

In recent years, \ml has emerged as a popular field of software development. \ml involves the use of statistical and computational techniques to develop predictive models from the data. The use of \ml has led to the development of several libraries and frameworks that simplify the development of \ml applications.

However, like any other software development project, \ml libraries are also prone to different issues and bugs \citep{islam2019comprehensive}. These problems can significantly impact the performance of \ml models, leading to incorrect predictions and results \citep{cheng2018manifesting}. Therefore, it is essential to identify, classify, and resolve bugs in \ml libraries as quickly as possible.

Despite the importance of issue-reporting systems in managing software bugs, there is limited empirical evidence on the performance of these systems in \ml libraries. This study aims to address this gap by conducting an empirical study of issue reporting systems in six popular \ml libraries. By analyzing issues reported in these libraries, we aim to provide insights into the most prevalent issues, the time taken to resolve them, and the effectiveness of issue reporting systems in managing the bug fixing process.

\begin{figure*}[ht]
\renewcommand\thefigure{I}
\includegraphics[width=\textwidth,height=8cm]{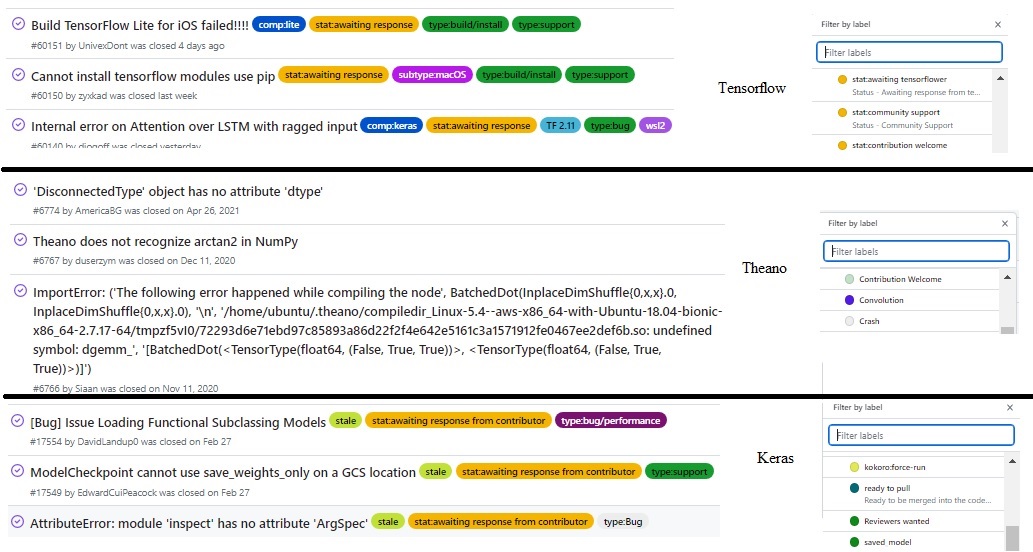}
\caption{Sample of \te, \ke, and \th issues and labels on \git}
\label{motivation}
\end{figure*}

In addition to managing software bugs through issue reporting systems, successful and efficient issue resolution requires the proper labeling of issue classification, categorization, type, and criticality. However, our manual study of reported issues in multiple \ml libraries, as depicted in \Cref{motivation}, revealed that many libraries lack such categorization. Furthermore, a majority of closed issues were left unupdated, either due to contributors or authors failing to respond, indicating a need for further investigation into their impact on issue resolution speed. We also found that many \ml libraries lack organization, with numerous dispersed labels that are difficult to comprehend. For instance, some issues in the \te library were assigned the status \textit{stat: awaiting tensorflower}, indicating their stage in the resolution process. Such a labeling technique can aid users and developers in determining which issues require attention, but it is not used in many other libraries, hampering issue resolution and the bug-fixing process.

Thus, it is crucial to examine the effects of these shortcomings on the bug-fixing process. The results of this study will be helpful to software developers, \ml researchers, and library maintainers in enhancing the dependability and quality of \ml libraries.

\section{Explanation and Definition of Concepts and terminology} \label{Sec::concepts}
In this section, we discuss the relevant concepts and terminologies contained in the dataset used in our study. We explain each concept in detail and discuss their implications for issue resolution and bug-fixing processes. We further explain the metrics adopted for determining the issue resolution effectiveness and bug fixing process of some \ml libraries
\subsection{Explanation and Definition of Concepts} \label{terminologies}
Several concepts were used in this study. These concepts are explained below and the others are defined in \Cref{definition}.
\subsubsection{\ml Library}
A \ml library is a collection of tools, algorithms, and functions that enables developers and users to build and deploy \ml applications. These libraries provide prebuilt components that can be used to create \ml models quickly and easily. Some of the most popular \ml libraries used as case studies in this experiment are \te, \ke, \sk, \py, \ca, and \th.
These libraries play a crucial role in the development of \ml applications. They provide developers with a platform to build and experiment with various models, algorithms, and techniques. However, these libraries are not free from bugs and issues, and hence, their maintenance is essential for ensuring smooth functioning.
\subsubsection{Latent Dirichlet Allocation (LDA)}
LDA is a statistical topic modeling technique that helps discover the underlying topics in a large corpus of text documents \citep{Blei2003}. LDA has many applications in natural language processing, such as document clustering, document classification, and information retrieval. It can also be used to analyze social media data, customer reviews, and other types of unstructured text data. One of the strengths of LDA is its ability to automatically discover the topics present in the corpus without the need for prior knowledge of the topics or manual labeling of documents. 
In this study, we used LDA to analyze the issue titles of the \ml libraries on \git. Even though LDA has some limitations, such as the sensitivity of the results to the number of topics specified and the quality of the initial word embeddings, it still serves the purpose of use in this study. Specifically, we extracted keywords that describe the issues reported by users to gain insight into the common problems faced by developers using these libraries. To achieve this, we used LDA tokenization capabilities. Tokenization is the process of splitting text into individual words or tokens. The text documents are first preprocessed by converting them into a collection of words or tokens and removing stop words, punctuation marks, and other non-relevant words. These collections were used to identify the relevant keywords used in this study. We choose this toolkit because it is a well established Natural Language Processing (NLP) tool that has been used in various prior research studies to analyze different information \citep{liang2014sentiment, li2015group, bundschus2009hierarchical, schwarz2018ldagibbs, panichella2013effectively}
\subsubsection{Statistical analysis}
This involves gathering and examining the data to uncover patterns and trends. Numerical analysis helps eliminate bias in the data evaluation. This method is highly beneficial for interpreting research findings, creating statistical models, and organizing surveys and studies \footnote{https://www.simplilearn.com/}. In this study, four statistical analyses were conducted: regression analysis, correlation analysis, percentage and frequency counts, and heatmap analysis.

Regression analysis is a widely used statistical technique for analyzing and modeling the relationship between a dependent variable and one or more independent variables. This helps to understand the factors that affect a particular outcome of interest. The main objective of the regression analysis is to identify the best-fitting mathematical model that describes the relationship between variables. For this study, an Ordinary Least Squares (OLS) model was used. OLS is a linear regression model that estimates the relationship between a dependent variable and one or more independent variables by minimizing the sum of the squared differences between the observed values of the dependent variable and the predicted values from the model. In this study, the OLS model was employed to determine how the number of comments on an issue affects the time taken to resolve the issue in \ml libraries.

A correlation analysis is a statistical measure that indicates the strength and direction of the relationship between two variables. This study aims to determine how the number of assignees/contributors to an issue affects the time it takes to close the issue in \ml libraries. The correlation can be positive, negative, or zero, indicating the presence or absence of a relationship between the variables. The correlation coefficient ranges from -1 to 1, with values closer to 1 indicating a stronger correlation.

Frequency counts and percentages were also used in this study. These measures are commonly used to summarize statistical data. The frequency count refers to the number of times a value occurs in a dataset, while the percentage is the frequency count, expressed as a proportion of the total number of observations. These measures are useful for summarizing categorical data, such as demographic variables, and can be used to identify patterns and trends in data. In this study, frequency count and percentage were used to determine the count and percentage of prioritized and unprioritized issues, percentage of issues in each library, and percentage of open and closed issues.

Heatmap analysis is a graphical representation of data in which different colors are used to represent different values. It is often used to visualize large datasets to identify patterns and trends that may be difficult to discern from the raw data. In this study, a heatmap was used to identify the different stages of each issue in each \ml library before closing. It has also been used to identify related issues mostly faced by \ml library users. These statistical methods were used in previous empirical studies \citep{bun2019ols, abd2020top, pak2020deep, budhwani2020creating}.

\begin{table}
\renewcommand\thetable{I}
\caption{\textit{Definition of Concepts}}
\label{definition}
    \begin{tabular}{|c|L|}
        \hline
        Concepts & \multicolumn{1}{c|}{\textbf{Description}}  \\
        \hline
        Issue ID	&  This concept refers to the identification number assigned to each issue report. \\
        \hline
        Issue state	& This refers to the current state of the issue reported in an issue-tracking system that indicate whether an issue has been acknowledged, assigned to someone, resolved, or closed.  \\
        \hline
        Issue title	& This is a brief summary of the issue that gives an idea of what the issue or bug is, but not so long as it becomes difficult to read.  \\
        \hline
        Issue body	& This refers to the details of the problem being reported, steps to reproduce the issue, any error messages, and other relevant information. \\
        \hline
        Labels	& This refers to keywords or tags used to categorize an issue or bug report. Labels can be used to indicate the type of issue or bug, its severity, or other relevant information. \\
        \hline
    \end{tabular}
\end{table}

\subsection{Bug fixing process metrics} \label{metrics}
In this study, we used the metrics explained in \Cref{metricsdef} to determine the effectiveness of issue resolution for bug fixing process in \ml libraries.
\begingroup
\renewcommand{\thetable}{II}
\begin{longtable}{|p{.20\textwidth} | p{.40\textwidth} | p{.40\textwidth} |} 
\caption{Metrics used in this study} 
\label{metricsdef}\\
\hline
Concept & \multicolumn{1}{c|}{\textbf{Description}} & \multicolumn{1}{c|}{\textbf{Implication}}  \\ 
\hline 
Issue Category	& Common issues or issue categories refer to the types of issues commonly reported by users and developers on the GitHub page of the library. These issues can be related to bugs, performance, documentation, or any other type of issues. & Common issues play a significant role in the bug fixing process, as they help developers identify the most pressing problems that need to be addressed. \\ \hline
Related Issue	&  The common  keywords extracted from issue titles, which were used to identify related issues and grouped into Performance, Validation, Algorithm, Testing, Compilation, Runtime, Miscellaneous, and Build for different \ml libraries, thus identifying the specific aspect of the library causing user problems. & By categorizing issues according to these categories, developers can easily identify and prioritize them, allowing them to focus on the most critical issues first and make more efficient use of their time and resources.\\
\hline
Issue Comments	& Issue comments refer to the total number of comments made on each issue by the developers, users, or contributors, and provide insights and suggestions on how to resolve the issue. & Issue comments are crucial in the bug fixing process, as they provide developers with feedback and insights on how the issue affects users and the community. \\
\hline
Issue Priority	& Issue priority refers to the priority level given to each issue. This priority is based on the severity of the issue, its impact on users and the community, and the complexity of the issue. & Issue priority plays a crucial role in the bug fixing process, as it helps developers prioritize the most critical issues first.\\
\hline
Issues Stage & This metric determines the current stage of an issue, track the progress of issues and to determine the next steps required for its resolution in bug-fixing process. & By tracking the issue status in \ml libraries, developers and library maintenance can quickly identify issues that require attention and allocate resources accordingly. \\
\hline
Issue contributors	& This refers to the people responsible for investigating an issue, proposing a solution, and implementing fixes when a new issue is opened. & Assigning and tracking issues to specific individuals ensures that tasks are being handled by someone who is accountable for them and that progress is being made. \\
\hline
Created date/time	& This refers to the date and time at which an issue or bug report is created. & This can help us to determine how effective and quick developers and library maintainers are in resolving issues in \ml libraries.\\
\hline
Closed date/time	& This refers to the date and time at which an issue or bug report is closed. 	& This can help us to determine how responsive developers and library maintainers are in resolving issues.\\
\hline
First response time	& This refers to the amount of time taken by developers or contributors to respond to an issue reported on the \ml library's GitHub page. & This can help the developers and contributors respond to issues as soon as possible to ensure that users are satisfied with the library and that they continue to use it.\\
\hline
Issue duration	& Issue duration, also known as time-to-close, refers to the time between when an issue is created and when it is finally closed or resolved. & The implication of this metric for the development team is that a short issue duration implies that they can quickly identify and resolve issues and close them in a timely manner.\\
\hline
\end{longtable}

\subsection{Issue keywords' mapping explanation} \label{group}
In this study, we extracted keywords from the titles of each generated issue reported in each \ml library. These keywords were used to explain the reporters’ main challenges, and how they are related with the issues category label. These keywords and how they are related to the bug fixing process are then grouped as follows in \Cref{grouptable}:

\begin{table}
\renewcommand\thetable{III}
\caption{\textit{Keyword groups}}
\label{grouptable}
    \begin{tabular}{|c|L|}
        \hline
        Group & \multicolumn{1}{c|}{\textbf{Description}}  \\
        \hline
        Compilation Issues	& These are related to problems in compiling the code. This can occur if there are missing dependencies, syntax errors, or other issues that prevent the code from being properly compiled. \\
        \hline
        Build Issues	& These issues are similar to compilation issues, but occur during the process of building software. This can include linking, packaging, or other steps involved in creating a distributable version of the software. \\
        \hline
        Runtime Issues	& These issues occur when software encounters problems while running. This can include crashes, errors, and unexpected behavior.  \\
        \hline
        Performance Issues	& These issues are related to the speed or efficiency of the software. This can include slow response times, high CPU usage, and other issues that affect the performance. \\
        \hline
        Testing Issues	& This type of issue can occur during software testing. This can include issues with test cases, test data, or the testing environment. \\
        \hline
        Validation Issues	& This type of issue is related to problems in ensuring that the software meets the desired specifications or requirements. This can include issues with the functionality, compatibility, or other aspects of the software that are important to the end user.\\
        \hline
        Algorithm Issues & Issues that can occur when there are problems with the underlying algorithms used in software. This can include issues with the accuracy, efficiency, or other aspects of the algorithm that impact its performance. \\
        \hline
        Functional Issues	& These type of issues are related to software functionality problems. This can include issues with specific features, user interfaces, and other aspects of software that impact its functionality. \\
        \hline
        Miscellaneous Issues	& These type of issues do not fit into any of the above mentioned categories. This can include issues with documentation, user experience, discussion, questions, or other aspects of the software that are important to the users. \\
        \hline
    \end{tabular}
\end{table}

\section{Methodology}
\label{Sec::methodology}
In this section, we elaborate on the experimental methodology employed in our study. Specifically, we discuss the dataset utilized, the data collection process, and the data analysis approach.

The framework of our study is depicted in \Cref{frame}. We developed a model to crawl issues from the \git repository of six \ml libraries. Subsequently, the extracted issues were imported into CSV files. To answer our research questions, we analyzed the issues and derived recommendations based on the findings. We also make the dataset and source code used in our experiment publicly available on our bitbucket repository \footnote{https://bitbucket.org/eyinlojuoluwa/issue-resolution/downloads/}.

\begin{figure*}[ht]
\renewcommand\thefigure{II}
\includegraphics[width=\textwidth,height=8cm]{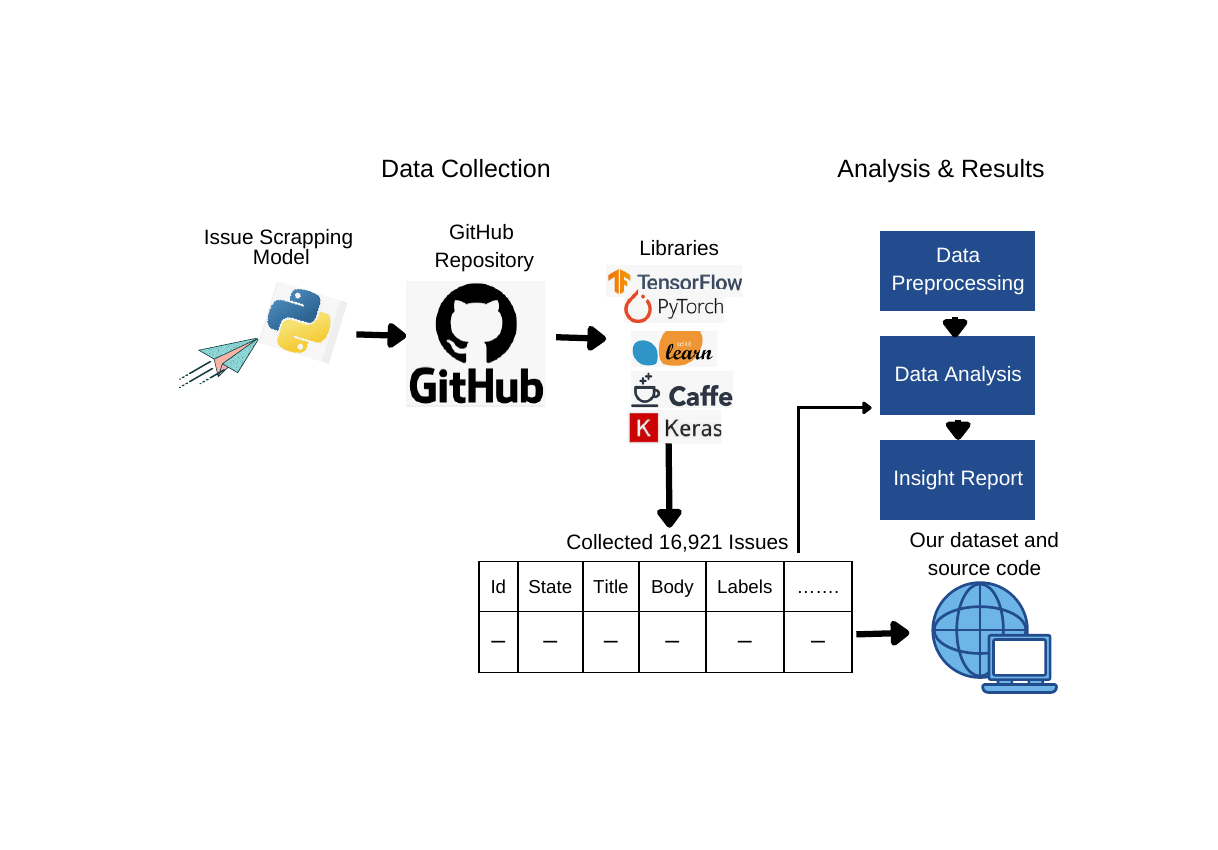}
\caption{An overview of our study design showing the process of data extraction and analysis}
\label{frame}
\end{figure*}

\noindent \textbf{Data Collection:} To gather our dataset, we utilized the \git rest API \footnote{https://docs.github.com/en/rest} to crawl reported issues within the studied \ml libraries. The \git rest API is a web-based platform that is commonly used for version control and collaboration, offering a web interface, desktop application, and a mobile application. This API offers endpoints for numerous types of operations, including managing repositories, issues, and pull requests, which can be accessed via HTTP requests and can be utilized in a variety of programming languages. As such, we leveraged this API to access and download all reported issues for six distinct \ml libraries on \git from February 28 2018 to February 28 2023.

We selected the following six \ml libraries: \te \footnote{https://github.com/tensorflow/tensorflow/issues}, \py \footnote{https://github.com/pytorch/pytorch/issues}, \ke \footnote{https://github.com/keras-team/keras/issues}, \sk \footnote{https://github.com/scikit-learn/scikit-learn/issues}, \th \footnote{https://github.com/Theano/Theano/issues}, and \ca \footnote{https://github.com/BVLC/caffe/issues}. These libraries were chosen because of their widespread use within the \ml community and their significant impact on the development of various \ml applications. By analyzing the reported issues within these libraries, we can gain insight into the common challenges and problems that developers face when working with these libraries. This information can inform improvements to library design and functionality, as well as aid in the development of new resources and tools to support \ml practitioners.

Each of the selected \ml libraries was introduced in a different year, with \te introduced in 2015, \py in 2016, \sk in 2007, \ke in 2015, \th in 2007, and \ca in 2013. We set the API to generate reported issues from five years, with the intention of generating issues for five years until February 28. As of March 4, we had generated these issues.

\setlength{\extrarowheight}{1pt}
\begin{table}
\renewcommand\thetable{IV}
\caption{\textit{Studied dataset}}
\label{dataset}
\small
\resizebox{\columnwidth}{!}{%
\begin{tabular}{lccccc}
\toprule
Library & Year & Extracted issue & Avg. issue/yr & Closed (\%) & Open (\%)\\
\midrule
\ca	&2013	&3396	&679.20	&74.53	&25.47\\
\ke	&2015	&2990	&598.00	&90.67	&9.23\\
\py	&2016	&2651	&530.20	&63.33	&36.67\\
\sk	&2007	&2699	&539.80	&74.10	&25.90\\
\te &2015	&2513	&502.60	&72.67	&27.33\\
\th	&2007	&2672	&534.40	&78.07	&21.93\\
\midrule
\textbf{Total}&&\textbf{16921}&\textbf{564.03}&\textbf{75.56}&\textbf{24.42}\\
\bottomrule
\end{tabular}}
\end{table}

The status of the generated issues is displayed in \Cref{dataset}. The first and second columns list the names and years in which each library was introduced, while the third and fourth columns show the total number of generated issues per library, as well as the average number of issues per year within the five-year timeframe. Columns five and six display the percentages of closed and open issues within the dataset, respectively. Across the six libraries, there were 16,921 reported issues, averaging 564 issues reported per year within the five-year timeframe. Of these, 75.56\% were closed, whereas 24.42\% remained open. For clarity, the average number of issues per year was calculated by dividing each extracted issue by the number of years of issues extracted. This implies that we divided the extracted issues by five.

\noindent\textbf{Data Processing and Analysis:} This step require us to extract the needed metrics as well as how they were extracted. In our study, for each of the generated issue, we extracted the metrics that were discussed in \Cref{metrics}. We took different dimension to process data for our research questions. We discuss how we extracted data and how we analyzed them for each research question.

\noindent\textit{Research Question-1:} In this RQ, we extracted the categories of all issues that were reported in each \ml library. However, the majority of the libraries studied did not have labels that could identify the category of the issue reported. For those that have labels for issue category, such as \te, \py, and \ke, in which they use "type" or "topic" to categorize the issues, we used these two labels to extract their issues categories. For the other libraries, we first used these two labels to extract issue categories. However, no results were obtained. Therefore, we  manually analyzed all labels in each issue and itemized what we assumed to be their respective categories. For \sk issues, we identified the following categories based on their issues: bugs, dependency, documentation, enhancement, meta-issues, build/CI, performance, validation, new features, regression, refractor, API. 
For the \th library, we identified crash, critical, enhancement, nice-to-have, optimization, regression, and stability. For the \ca library, we identified bugs, compatibility, enhancement, parallelism, testing, opencl, speed-up, questions, documentation, community, build, and installation. We then used these labels to represent possible issue categories. We understand that this might not be the best method to determine the main issue faced by the reporter. However, we ensured that this technique was repeated many times by two different authors to avoid omitting other labels. This is one of the challenges faced by libraries that lack proper labelling. They are required for good organization and labelling of each issue. \Cref{issue_extra} displays the overall steps of achieving this goal.
Therefore, frequency count and percentage were used to analyze the results and then visualized with a bar chart.

\begin{algorithm}[H]
\caption{Issues Category Extraction}\label{issue_extra}
\begin{algorithmic}
\Procedure{ExtractIssueCategories}{$libraries$}
\ForAll{$library \in libraries$}
\State $categories \gets \varnothing$
\If{$library$ has "type" or "topic" labels}
\State $categories \gets$ extract categories using the labels
\Else
\State manually analyze and itemize issue categories for $library$
\State run algorithm again on $library$ with updated categories
\EndIf
\State $issue_categories[library] \gets categories$
\EndFor
\State \textbf{return} $issue_categories$
\EndProcedure
\end{algorithmic}
\end{algorithm}

Where:
\begin{itemize}
\item $libraries$ is a list of machine-learning libraries
\item $issue_categories$ is a dictionary with libraries as keys and a list of categories as values
\end{itemize}

\noindent\textit{Research Question-2:} To answer the research question, we utilized natural language processing techniques to analyze all reported issues and extracted the main issues faced by users and developers from each issue's title. Initially, we analyzed the issues' body, but due to the presence of noisy information and the lack of informative keywords, we opted to use Latent Dirichlet Allocation (LDA) from \sk \footnote{https://scikit-learn.org/stable/modules/generated/sklearn.decomposition.LatentDirichletAllocation.html} to identify the most important keywords within each title of each issue. The goal of LDA is to identify the most important topics or keywords within text data. To achieve this, we created a Term Frequency-Inverse Document Frequency (TF-IDF) vectorizer object to extract keywords that explain the issues reported by developers and users. TF-IDF is a common weighting scheme in natural language processing that reflects the importance of a keyword to a particular issue title. To limit the size of the resulting feature matrix, we set the maximum number of features to 1000.

Next, each issue's title was transformed into a TF-IDF matrix using the vectorizer, and we used the LDA to model the 10 topics and fit them to the TF-IDF matrix. We calculated the topic distribution for each issue using the LDA model, identified the topic with the highest probability, and extracted the top two keywords for each topic from the feature matrix. We limited the number of extracted keywords to two to combine the first and second keywords and create a single combined word. To filter out uninformative keywords, we removed keywords with fewer than three characters, such as \textit{np} and \textit{tf}, which often represent library names. Additionally, we filtered out the name of the library that we were working on for each issue. For example, if we were generating keywords from the \textit{\te} library, we filtered out the word \textit{\te} from the keywords.

The resulting keywords were concatenated into a single string and classified into various categories, including performance, validation, algorithm, testing, compilation, runtime, build, and miscellaneous categories. This classification helped us identify the types of problems faced by the users. For instance, if the keyword is \textit{libtorch failed}, we classified it as a compilation issue. \Cref{key} shows the overall steps taken to achieve this aim. We then used heatmap to visualize the results. \Cref{Sec::concepts} provides more information about LDA. 

\begin{algorithm}
\caption{Keyword Extraction}\label{key}
\begin{algorithmic}
\State Set maximum number of features to $1000$
\State Instantiate a TF-IDF vectorizer object from the Latent Dirichlet Allocation
\State Fit the vectorizer object to the titles of GitHub issues from each machine learning library repository
\State Transform each issue's title into a TF-IDF matrix
\State Instantiate an LDA model with $10$ topics and fit it to the TF-IDF matrix
\For{each issue title $j$}
\State Calculate the topic distribution using the LDA model and store in $P_j = (p_{j,1}, p_{j,2},...,p_{j,10})$
\State Identify the topic with the highest probability for the issue, i.e., $\arg\max_{i\in{1,2,...,10}}p_{j,i}$
\State Remove keywords with fewer than three characters, i.e., $|C_j| < 3$
\State Filter out the name of the library being worked on for each issue
\If{$\text{library name}\in C_j$}
\State Remove $\text{library name}$ from $C_j$
\EndIf
\State Extract the top two keywords for each topic from the feature matrix and store in $K_j = (k_{j,1}, k_{j,2})$
\State Combine the first and second keywords into a single combined word and store in $C_j$
\State Concatenate the resulting keywords into a single string and store in $S_j$
\State Classify the concatenated keywords into performance, validation, algorithm, testing, compilation, runtime, build, and miscellaneous categories and store in $C_{j}'$
\EndFor
\end{algorithmic}
\end{algorithm}

Note in \Cref{key} that $p_{j,i}$ is used to represent the probability of issue $j$ belonging to topic $i$, $k_{j,m}$ to represent the $m$-th keyword extracted for issue $j$, $C_j$ to represent the combined word for issue $j$, $S_j$ to represent the concatenated keywords for issue $j$, and $C_{j}'$ to represent the category for issue $j$.

\noindent\textit{Research Question-3:} To answer this research question, we performed a simple linear regression analysis on all the issues using the statsmodels \footnote{https://www.statsmodels.org/stable/index.html} library in Python to find the relationship between the duration of issues and the total number of comments. We used the \textit{total number of comments} as the independent variable, and the \textit{duration of the issue} as the dependent variable. We then used the \textit{fit()} method to estimate regression coefficients and other statistical information. The regression coefficients, standard errors, t-statistics, p-values, and R-squared values are reported. This process was repeated for each \ml library.

\noindent\textit{Research Question-4:} To answer this question, we utilized the \textit{date of creation} and \textit{closing dates} of issues to assess their resolution times. To achieve this goal, we analyzed both open and closed issues, examining the duration an issue remained unaddressed before receiving any response, and the overall duration it remained open until its closing date.

To calculate the length of time that an issue spent waiting for its first response, we employed a straightforward approach. For both open and closed issues, we subtracted the issue's creation date from the date on which the response was posted. This allowed us to determine the time required for the first response to be updated for each issue. In instances where an issue remained unaddressed and was subsequently closed, we used the time that the issue remained unaddressed as the total duration.

In cases where an issue remained open but was yet to receive a response, we defined the first response time as the difference between the issue's creation date and the date on which the issues were downloaded from the repositories. This helped us understand how long it took for the issues to be acknowledged and addressed.

Furthermore, we calculated the total duration of an issue, which was defined as the time between its creation date and closing date for closed issues. For open issues, the total duration was calculated by subtracting the issue creation date from the date it was downloaded from the repositories.

As a result of these calculations, we were able to determine two different average response times for each issue: the first response time and the total duration. Both measures are expressed in minutes, allowing us to make precise and reliable comparisons between different issues.

\noindent\textit{Research Question-5:} To address this research question, we extracted priority labels for each issue from the studied \ml libraries. We limited our focus to closed issues, because the duration of open issues could not be precisely determined until they were resolved. By examining closed issues, we obtained a more precise and comprehensive understanding of how issue priority influences the duration of an issue within \ml libraries. We then calculated the percentages of \textit{prioritized} and \textit{unprioritized} issues according to their respective durations.

\noindent\textit{Research Question-6:} To answer this research question, we analyzed the closed issues and extracted items with the "stat" label from the studied \ml libraries. "Stat" is a label used to track the status of an issue or pull request, indicating whether it is actively being worked on, whether a fix has been proposed or implemented, or whether it is waiting for additional information or action from the reporter or another contributor. However, only \te and \ke have this label; therefore, we manually listed labels that might likely represent the status of issues in other \ml libraries.

During this manual analysis, we discovered that there was no specific label in \py that indicated the state or status of the issues reported in the library. Consequently, it may be more challenging to quickly identify the status of the issue reported in \py. However, we were able to identify specific status labels for each of the following libraries: \ca (\textit{abandoned}, \textit{in progress}, \textit{merge pending}, \textit{ready for review}, \textit{unfixable-issue}, \textit{duplicate}), \th (\textit{CCW - in progress}, \textit{CCW - reserved}, \textit{CCW}, \textit{closed as we stop development}, \textit{contribution welcome}, \textit{not the place to request support}), and \sk (\textit{waiting for a reviewer}, \textit{waiting for a second reviewer}, \textit{contribution welcome}).

Furthermore, there were no labels indicating whether an issue was resolved in the issue repositories. Therefore, we utilized the \textit{stalled} or \textit{stale} label. \textit{Stalled} or \textit{stale} typically refers to an issue that has not seen any recent activity or progress. A \textit{stalled} or \textit{stale} issue may have been opened a long time ago but has not seen any updates or changes, or it may have been assigned to a contributor who has not made any progress on it for an extended period.

In this case, we assumed that if an issue is closed and has no \textit{stalled} or \textit{stale} label, such issue was resolved, even though it does not necessarily guarantee that the issue was resolved satisfactorily or that it is no longer relevant. Similarly, we assumed that if an issue is closed and has either \textit{stalled} or \textit{stale}, such issue was not resolved before closing, even though it does not necessarily mean that the issue was not resolved satisfactorily or that it is still relevant.

We also assumed that if an issue is still open and does not have \textit{stalled} or \textit{stale}, such an issue awaits a response. Similarly, we assumed that if an issue is open and has \textit{stalled} or \textit{stale}, such an issue has been abandoned. We use these four assumptions to determine whether an issue was \textit{resolved}, \textit{awaiting a response}, \textit{unresolved}, or \textit{abandoned}. We included these assumptions to determine the state of the generated issues. This information, along with the duration of the issues, was used to answer the research question and only closed issues were considered. We then used a heatmap from matplotlib\footnote{https://matplotlib.org/3.5.3/api/\_as\_gen/matplotlib.pyplot.html} in python to visualize the results.

\noindent\textit{Research Question-7:} To address this research question, we computed the length of the assignee label for each closed issue, to derive the total number of contributors involved. Furthermore, we determined the duration of each issue as previously outlined. To establish the association between the number of assignees or contributors and the duration of an issue, we utilized the groupby method in Python to correlate the metrics, followed by the use of the corr() method from the pandas library, which computes the correlation coefficient between two variables \footnote{https://pypi.org/project/pandas/}. Finally, we present the correlation coefficients between the number of contributors and the duration of issues using a bar chart.

\section{Study Results}
\label{Sec::result}
This section presents the findings of our empirical investigation. For each RQ, we present the results and their implications for the issue resolution and bug-fixing process.  
\subsection{\RQOne}
In this research question, we reported only the top five categories for each \ml library instead of all categories. We report the result in \Cref{RQ1} and visualize the result in \Cref{RQ1visual}. The top 1 categorized issues are represented with \textit{crimson} color, top 2 with \textit{orange} color, top 3 with \textit{gold}, top 4 with \textit{limegreen}, and top 5 with \textit{deepskyblue} color. \\
\textbf{Findings:} \Cref{RQ1} show that the categorized issues in the \ca library are Question, OpenCL, Bug, and Documentation. Of the thousands of issues reported in the \py library, only one is categorized as not user-facing, performance, improvements, and build. Contrary to what was observed in the \py library, the majority of issues in the \th library were categorized as optimization, crash, enhancement, and bug. Similarly, the majority of issues are also classified under Bug, New Feature request, Documentation, and Build/CI in the \sk library, Support, Bug/Performance, feature, documentation in the \ke library, and Bug, Build/Install, Support, feature, and performance in the \te library.

\begin{figure*}[ht]
\renewcommand\thefigure{III}
\includegraphics[width=\textwidth,height=8cm]{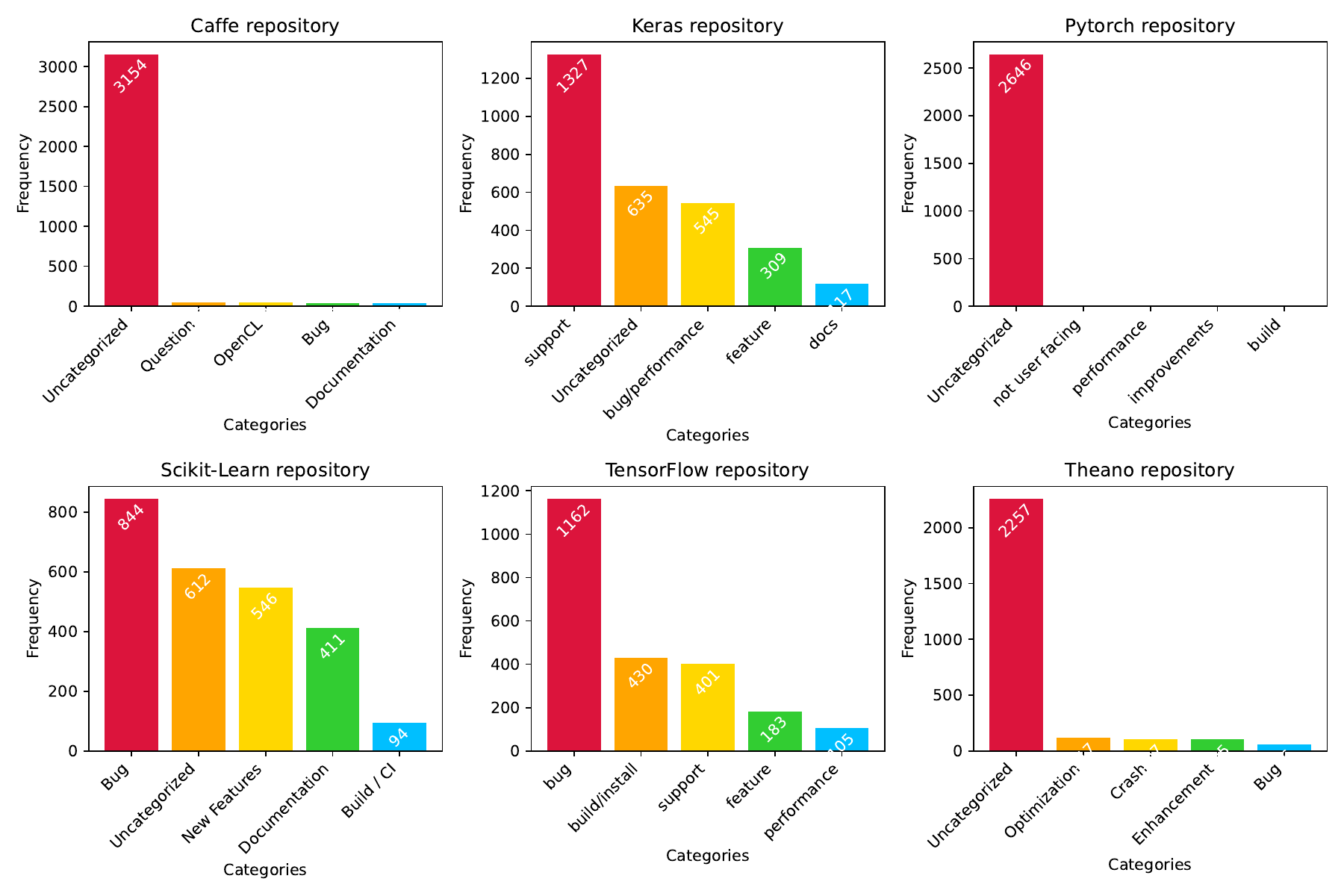}
\caption{Top 5 issue categories of each \ml}
\label{RQ1visual}
\end{figure*}

\setlength{\extrarowheight}{1pt}
\begin{table}
\renewcommand\thetable{V}
\caption{\textit{Reported Issues category}}
\label{RQ1}
\small
\resizebox{\columnwidth}{!}{%
\begin{tabular}{ccc|ccc|ccc}
\toprule
\multicolumn{3}{c}{\ca} & \multicolumn{3}{c}{\py} & \multicolumn{3}{c}{\th}\\
Category & count & \% & Category & count & \% & Category & count & \% \\ 
\midrule
Uncategorized	& 3154	&92.87	&Uncategorized	&2646	&99.81	&Uncategorized	&2257	&84.47\\
Question	&46	&1.35	&Not user facing	&1	&0.04	&Optimization	&117	&4.38\\
OpenCL	&44	&1.3	&Performance	&1	&0.04	&Crash	&107	&4.00\\
Bug	&42	&1.24	&Improvements	&1	&0.04	&Enhancement	&105	&3.93\\
Documentation	&34	&1.00	&Build	&1	&0.04	&Bug	&56	&2.10\\
\bottomrule
& & & & & & & &\\

\multicolumn{3}{c}{\sk} & \multicolumn{3}{c}{\ke} & \multicolumn{3}{c}{\te}\\
Category & count & \% & Category & count & \% & Category & count & \% \\
\midrule
Bug	&844	&31.27	&support	&1327	&44.38	&bug	&1162	&46.24\\
Uncategorized	&612	&22.68	&Uncategorized	&635	&21.24	&build/Install	&430	&17.11\\
New Features	&546	&20.23	&Bug/Performance	&545	&18.23	&Pupport	&401	&15.96\\
Documentation	&411	&15.23	&Feature	&309	&10.33	&Feature	&183	&7.28\\
Build / CI	&94	&3.48	&Documentation	&117	&3.91	&Performance	&105	&4.18\\

\bottomrule
\end{tabular}}
\end{table}

\Cref{RQ1visual} shows that a significant number of issues reported in \ml libraries are not categorized, as it takes the first or second position in five libraries: \ca, \py, \th, \sk, and \ke. This could limit the ability of library maintainers to identify the most critical issues that require attention. However, categorized issues provide insight into the most common problems faced by users of \ml libraries.

The most commonly categorized issue across the libraries was "Bug," as it appears in five \ml libraries: \ca, \th, \sk, \ke, and \te, suggesting that developers need to focus on improving the quality of their libraries. The "Documentation" category was also common, as it also appears in three of the \ml libraries, \ca, \sk, and \ke, highlighting the need for developers to ensure that their documentation is up-to-date, comprehensive, and easy to understand.

Furthermore, several issues are common across multiple libraries. For example, "performance" was a common issue in \py, \ke, and \te. This suggests that developers must focus on optimizing their libraries to improve their performance.

\noindent \textbf{Implication:} The implication of this study to issues resolution and bug fixing process is that library maintainers and developers should focus on improving the quality of their libraries to reduce the number of bugs reported. They should prioritize fixing critical bugs and improving documentation to reduce the number of "uncategorized" issues. Developers can also benefit from user engagement to gather feedback, understand the issues users face, and prioritize bug fixes accordingly.

\noindent\textbf{Summary:} The majority of issues in the studied \ml libraries were not categorized. However, the categorized issues provide insight into the most common problems faced by users of \ml libraries. Developers should take note of these issues, prioritize fixing critical bugs, and improve their documentation to ensure that their libraries are of high quality.

\subsection{\RQTwo}
\noindent\textbf{Findings:} \Cref{RQ2} present the top five keywords that elucidate the challenges encountered by the users of the six \ml libraries, emphasizing on both open and closed issues.  \ca library users frequently mention the following keywords in their reported issues; failed checks, compilation, layers, files, and make-errors. \py library users mention torch onnx, torchscript, backend building, libtorch failures, and 'onnx models. In the case of the \th library, users mentioned errors, scan tests, optimization, tests, and support. On the other hand, the most common keywords extracted from the reported issues in \sk library include failed operations, sparse matrices, missing values, failing \sk functions, and k-means. \ke library users face issues related to models, model loading, custom features, layer attributes, and import validation, while \te library users often mention build delegates, training faults, tflite models, memory usage, and running Python code.

\setlength{\extrarowheight}{1pt}
\begin{table}
\renewcommand\thetable{VI}
\caption{\textit{Related Issues category}}
\label{RQ2}
\small
\resizebox{\columnwidth}{!}{%
\begin{tabular}{ccc|ccc|ccc}
\toprule
\multicolumn{3}{c}{\ca} & \multicolumn{3}{c}{\py} & \multicolumn{3}{c}{\th}\\
Keyword & count & \% & Keyword & count & \% & Keyword & count & \% \\ 
\midrule
Failed check	&419	&12.34	&Torch onnx	&327	&12.33	&error	&320	&11.98\\
Compile	&413	&12.16	&Torchscript	&319	&12.03	&Scan test	&304	11.38\\
Layer	&365	&10.75	&Build backend	&292	&11.01	&Optimization	&292	&10.93\\
File	&362	&10.66	&Libtorch failed	&282	&10.64	&Tests	&291	&10.89\\
Make error	&353	&10.39	&Onnx model	&273	&10.3	&Support	&285	&10.67\\
\bottomrule
& & & & & & & &\\

\multicolumn{3}{c}{\sk} & \multicolumn{3}{c}{\ke} & \multicolumn{3}{c}{\te}\\
Keyword & count & \% & Keyword & count & \% & Keyword & count & \% \\
\midrule
Failed	&302	&11.19	&Model	&423	&14.15	&Build delegate	&292	&11.62\\
Matrix sparse	&301	&11.15	&Model loading	&358	&11.97	&Training fault	&290	&11.54\\
Missing values	&293	&10.86	&Custom	&341	&11.4	&Tflite model 	&277	&11.02\\
Sklearn failing	&292	&10.82	&Layer feature	&326	&10.9	&Memory	&248	&9.87\\
Kmeans	&286	&10.6	&Import validation 	&287	&9.6	&Running python	&246	&9.79\\

\bottomrule
\end{tabular}}
\end{table}

These keywords were grouped into categories and then termed as related issues, including Performance, Validation, Algorithm, Testing, Compilation, Runtime, Miscellaneous, and Build. \Cref{RQ2visual} illustrates the relationship between the related issues that users face, indicating that Testing and Runtime issues are the most common challenges in all six \ml libraries. Specifically, the libraries \ca, \sk, \th, and \ke have issues related to testing, whereas \te, Pytorch, \ke, and \ke have issues related to runtime. Additionally, both the \ca and \ke libraries encountered issues in both the Runtime and Testing categories.

\begin{figure*}[ht]
\renewcommand\thefigure{IV}
\includegraphics[width=\textwidth,height=8cm]{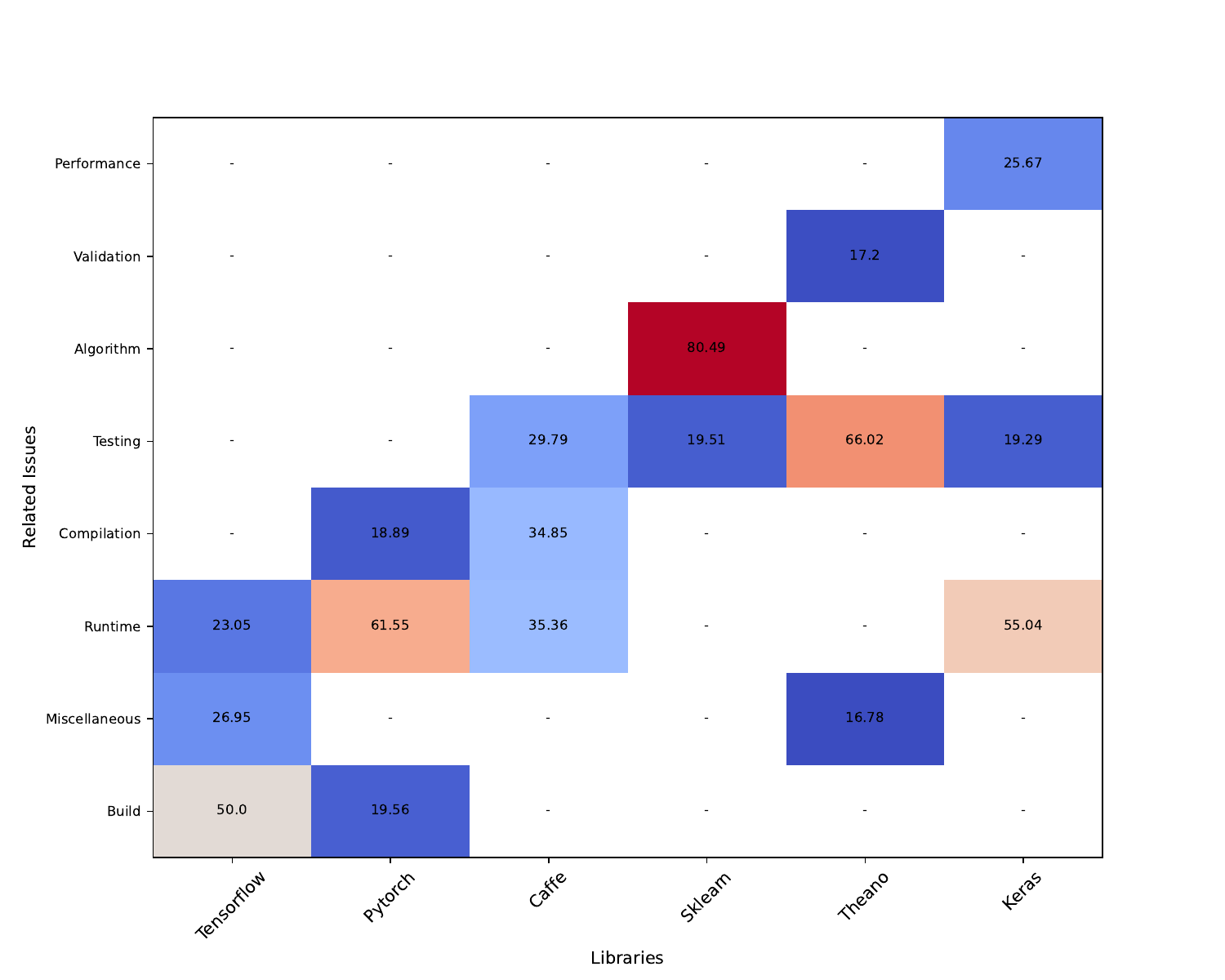}
\caption{Related Issue Categories}
\label{RQ2visual}
\end{figure*}

To answer this research question, the related issue that is mostly faced by \ml library users is a combination of Testing and Runtime issues. This is evident from \Cref{RQ2visual}, which illustrates the relationship between the related issues faced by users in all six \ml libraries. Specifically, four libraries (\ca, \sk, \th, and \ke) have issues related to Testing, while four libraries (\te, \py, \ca, and \ke) have issues related to Runtime.

\noindent\textbf{Implication:} These results have important implications for library maintainers and bug-fixing process. By identifying the most common issues faced by users, library maintainers can prioritize development efforts and allocate resources more effectively. For example, if a library suffers from many runtime issues, then library maintainers might focus on improving the speed and efficiency of the code. Similarly, if a library is experiencing many testing issues, then library maintainers might focus on improving the testing infrastructure and tools to make it easier for users to identify and fix bugs. In addition, this knowledge provides insights into how library maintainers can improve their processes and develop more user-friendly software.

\subsection{\RQThree}
\noindent\textbf{Findings: }The OLS regression results in table shows the estimated coefficients, standard errors, t-statistics, p-values, and 95\% confidence intervals for a linear regression model that relates the duration of issues (measured in days) to the total number of comments on the issue for six \ml libraries: \ca, \ke, \py, \sk, \te, and \th. The intercept represents the estimated duration of an issue when the total number of comments was zero.

\setlength{\extrarowheight}{1pt}
\begin{table}
\renewcommand\thetable{VII}
\caption{\textit{OLS Regression Results of duration of issues and total number of comments}}
\label{RQ3}
\small
\resizebox{\columnwidth}{!}{%
\begin{tabular}{ccccccc}
\toprule
Library & total duration & Coefficient & Standard Error & t & $P>|t|$ & R-squared\\
\midrule
\multirow{2}{*}{\ca} & Intercept (constant term)	&7.47E+05	&2.35E+04	&31.84	&0.000	&\multirow{2}{*}{0.001}\\
&total number of comment	&7680.3842	&3497.985	&2.196	&0.028&\\
&&&&&&\\
\multirow{2}{*}{\ke} & Intercept (constant term)	&4.69E+05	&9672.328	&48.473	&0.000	&\multirow{2}{*}{0.017}\\
&total number of comment	&-1.14E+04	&1593.09	&-7.152	&0.000&\\
&&&&&&\\
\multirow{2}{*}{\py} & Intercept (constant term)	&3.63E+05	&1.30E+04	&27.796	&0.000	&\multirow{2}{*}{0.005}\\
&total number of comment	&8359.6663	&2353.822	&3.552	&0.000&\\
&&&&&&\\
\multirow{2}{*}{\sk} & Intercept (constant term)	&2.42E+05	&9247.597	&26.15	&0.000	&\multirow{2}{*}{0.011}\\
&total number of comment	&5318.7783	&990.426	&5.37	&0.000&\\
&&&&&&\\
\multirow{2}{*}{\te} & Intercept (constant term)	&7.28E+04	&3908.277	&18.625	&0.000	&\multirow{2}{*}{0.015}\\
&total number of comment	&3045.0453	&498.619	&6.107	&0.000&\\
&&&&&&\\
\multirow{2}{*}{\th} & Intercept (constant term)	&1.03E+06	&3.73E+04	&27.643	&0.000	&\multirow{2}{*}{0.004}\\
&total number of comment	&-1.79E+04	&5312.036	&-3.364	&0.001&\\

\bottomrule
\end{tabular}}
\end{table}

Based on the result in \Cref{RQ3}, we make the following observations:
The coefficient of the total number of comments is positive for all libraries, except for \ke and \th, where it is negative. This suggests that, as the number of comments on an issue increases, the duration of the issue also tends to increase, except for \ke and \th, where more comments are associated with shorter duration of issues.

The coefficient of the total number of comments was statistically significant (p-value < 0.05) for all libraries except for \ca, where the p-value was close to 0.05. This suggests that the total number of comments has a significant effect on the duration of issues for all libraries, except for \ca, where the effect is not as strong.

The R-squared values for the six models were low, ranging from 0.001 to 0.017. This suggests that the total number of comments alone cannot explain much of the variation in the duration of the issues for these libraries.
Similarly, we cannot make a definitive conclusion based on this result alone. However, we observed that \te has the smallest intercept (constant term) among the six libraries, which suggests that on average, issues are resolved more quickly in \te than in the other libraries. However, there might be other factors that affect the duration of the issues besides the total number of comments.

The regression results further suggest that libraries with larger coefficients for the total number of comments tend to take longer to resolve issues when there are more comments. This may imply that these libraries have a larger user base and, therefore, receive more feedback and bug reports, which takes longer to process. In this case, software issues must be solved quickly so that users can receive bug fixes and new features faster, which can improve user experience and productivity. In addition, fast bug fixes can help to maintain the reputation of a library and attract more users.

To answer RQ3, we can see that the number of comments on an issue has a significant impact on the time it takes to resolve the issue for all six \ml libraries. The negative coefficients for the \ke and \th libraries indicate that as the number of comments on an issue increases, the time taken to resolve the issue decreases, while the positive coefficients for the other four libraries suggest that the time to resolve issues increases with an increase in the number of comments.

\noindent \textbf{Implication:} This result is important for the issue resolutions and bug fixing process, as it suggests that monitoring the total number of comments on issues can provide insights into the duration of issues in \ml libraries. Furthermore, identifying factors that affect the duration of issues can help developers prioritize bug-fixing efforts and improve the efficiency of the bug-fixing process. 

\subsection{\RQFour}
\noindent \textbf{Findings:} The table below displays the average response times for issues reported in six different \ml libraries: \ca, \ke, \py, \sk, \te, and \th. This research question is important because it helps to establish user expectations and identify opportunities for improving response times.

\setlength{\extrarowheight}{1pt}
\begin{table}
\renewcommand\thetable{VIII}
\caption{\textit{Average response time of the reported issues}}
\label{RQ4}
\small
\resizebox{\columnwidth}{!}{%
\begin{tabular}{ccccccccccc}
\toprule
Library & Total Issue & Avg. Issue/yr. & Closed (\%) & Open (\%) &FR (mean) &FR (days) & WR (mean) & WR (days) & TR (mean) & TR (days) \\
\midrule
\ca	&3396	&339.60	&74.53	&25.47	&51279.67	&35.61	&273564.27	&189.98	&770098.48	&534.79\\
\ke	&2990	&373.75	&90.67	&9.23	&87363.57	&60.67	&90005.67	&62.50	&427829.67	&297.10\\
\py	&2651	&378.71	&63.33	&36.67	&23759.77	&16.50	&109626.02	&76.13	&388664.45	&269.91\\
\sk	&2699	&168.69	&74.10	&25.90	&13015.91	&9.04	&33692.38	&23.40	&266306.53	&184.94\\
\te	&2513	&314.12	&72.67	&27.33	&3349.73	&2.33	&7499.03	&5.21	&91067.72	&63.24\\
\th	&2672	&167.00	&78.07	&21.93	&67848.59	&47.12	&308666.42	&214.35	&954883.17	&663.11\\
\bottomrule
\end{tabular}}
\end{table}

\Cref{RQ4} shows the total number of issues extracted for each library, the average number of issues reported in each library's repositories per year, the percentage of closed and open issues for each library, the average time that each issue spent before receiving any response from contributors in minutes and days (\textit{First Response (FR)}), the average time that an issue spent without receiving any response in minutes and days (\textit{Without Response (WR)}), and the total lifespan of an issue in minutes and days (\textit{Total Response (TR)}). The analysis includes both closed and open issues and reveals that \ke has the highest number of closed issues, whereas \py has the highest number of open issues at the time of the experiment.

\Cref{RQ4} also reveals that \te and \sk are highly responsive to user feedback, with the lowest times for receiving the first response, at most 2 and 9 days, respectively. Similarly, both libraries also had the shortest average times for which issues remained without a response at 5 and 23 days, respectively. This suggests that development teams actively monitor the repositories and are committed to promptly addressing user concerns. This level of responsiveness indicates that \te and \sk development teams value user feedback and are dedicated to ensuring that the libraries are as reliable and user-friendly as possible.

Moreover, the table shows that \te and \sk libraries have a significantly lower average resolution time than the other libraries, with 63.24 and 184.94 days, respectively. This result suggests that the development communities working on these libraries are particularly efficient in resolving issues. It is possible that these communities have more resources and expertise available for troubleshooting and resolving issues, or that the libraries are designed in such a way as to make it easier to address issues quickly.

The result also supports the findings from RQ1, which shows that only \te, \sk, and \ke have clear categories of issues, such as bugs and support. Having a well-defined category label can help developers and maintainers quickly identify and prioritize issues and direct them to appropriate experts, who can resolve them more efficiently.

Finally, the results from RQ2 show that \te users are mostly faced with build-related issues, whereas \sk library users face algorithm-related issues. In the context of this result, it can be concluded that build- and algorithm-related bugs are likely to be attended to early for possible early resolution.

To answer RQ4, the average response time varies from library to library, depending on many factors. However, \te and \sk have the shortest response times, which makes the libraries resolve issues more quickly than the other libraries. This demonstrates the importance of responsiveness and efficient issue resolution in \ml libraries. By addressing issues promptly and efficiently, developers can prevent issues from escalating and causing further problems, which can save time and effort in the long run.

\subsection{\RQFive}
\noindent\textbf{Findings:} \Cref{RQ5} provides an overview of prioritized and unprioritized issues for five popular \ml libraries: \te, \py, \ca, \sk, and \th. For prioritized issues, the table lists the percentage of issues categorized as high, medium, or low priority; the total number of comments received for these issues; the rate at which the library received comments; and the average time and days taken to close these issues. The rate at which an issue received a comment was calculated as the percentage of the total comments over the percentage of the total prioritized or unprioritized issues. For unprioritized issues, the table shows the percentage of issues that were not prioritized, total number of comments received for these issues, rate at which the library received comments, and average time and days taken to close these issues.

\setlength{\extrarowheight}{1pt}
\begin{table}
\renewcommand\thetable{IX}
\caption{\textit{Issues Prioritization}}
\label{RQ5}
\small
\resizebox{\columnwidth}{!}{%
\begin{tabular}{ccccccccc}
\toprule
\multicolumn{9}{c}{Prioritized Issues}\\
Project & High (\%) & Medium (\%) & Low (\%) &$\sum$Comment & Comment (\%) & Rate & Avg. Time(minutes) & Avg. Time(days)\\ 
\midrule
\te	&0	&0	&0	&0	&0	&0.00	&0	&0\\
\py	&5.21	&0.00	&0.11	&751	&13.83	&2.60	&179949.97	&124.97\\
\ca	&0	&0	&0	&0	&0	&0.00	&0	&0\\
\sk	&0.19	&0.00	&0.07	&66	&0.76	&2.92	&168422.85	&116.96\\
\th	&1.01	&0.34	&0.52	&237	&2.50	&1.34	&482668.19	&335.19\\
\midrule
\multicolumn{9}{c}{Unprioritized Issues}\\
Project & \multicolumn{3}{c}{No-priority (\%)} &$\sum$Comment & Comment (\%) & Rate & Avg. Time(minutes) & Avg. Time(days)\\
\te	&\multicolumn{3}{c}{72.7}	&11212	&100	&1.38	&37312.71	&25.91\\
\py	&\multicolumn{3}{c}{58.02}	&4678	&86.17	&1.49	&117464.38	&81.57\\
\ca	&\multicolumn{3}{c}{74.53}	&7819	&100	&1.34	&113375.82	&78.73\\
\sk	&\multicolumn{3}{c}{73.84}	&8669	&99.24	&1.34	&90238.62	&62.66\\
\th	&\multicolumn{3}{c}{76.2}	&9254	&97.50	&1.28	&205254.66	&142.54\\

\bottomrule
\end{tabular}}
\end{table}

It can be observed from \Cref{RQ5} that \te and \ca did not prioritize any closed issue in their libraries. However, \py, \sk, and \th prioritized some issues, with \py having the highest percentage of high-priority issues (5.21\%). The rate at which the prioritized issues received comments was higher than that of the unprioritized issues, which means that prioritized issues took longer to close than unprioritized issues. This could be attributed to the fact that prioritized issues receive more engagement from developers, resulting in more discussion and deliberations before a solution is agreed upon. However, unprioritized issues tend to be straightforward and require less attention from developers, leading to faster resolution.

To answer this research question, issue prioritization plays a significant role in the duration of issue resolution. While prioritizing critical issues is essential, it is equally important to address other issues efficiently to maintain the health of a software project.

\noindent \textbf{Implication:} The implication of this result on the development of \ml libraries is that issue prioritization can significantly impact the duration of issue resolution. Although prioritizing issues can ensure that critical bugs or features are addressed promptly, it can also cause delays in resolving other issues. Therefore, it is essential to strike a balance between prioritizing critical issues and addressing other issues in a timely manner.
\ml developers can adopt strategies to minimize the time spent on prioritized issues. First, they can ensure that the issues are thoroughly analyzed and discussed before being prioritized. This can help identify any potential roadblocks and enable developers to better prepare for resolving the issue. Second, developers can delegate issues to other contributors or community members who have the necessary skills and experience to tackle the issue. This can help distribute the workload and accelerate the issue resolution. Finally, developers can strive to improve their communication and collaboration processes to ensure that the issue resolution is performed efficiently without compromising the quality of the solution.
\subsection{\RQSix}
\textbf{Findings:} This research question assessed the status of each closed issue in each \ml library. Of the six studied libraries, the \py library issue repository does not have any labels that can identify the status of the issues. Therefore, this library was excluded from the analysis.

\Cref{RQ6visual1} illustrates the top 5 ranked issue statuses for each \ml library, with each bar representing the percentage of issues for each status. It is evident from the visualization that the status of the issues in the \sk library can only be resolved or unresolved, and no other status is available, as in the case of other libraries.

\begin{figure*}[ht]
\renewcommand\thefigure{V}
\includegraphics[width=\textwidth,height=8cm]{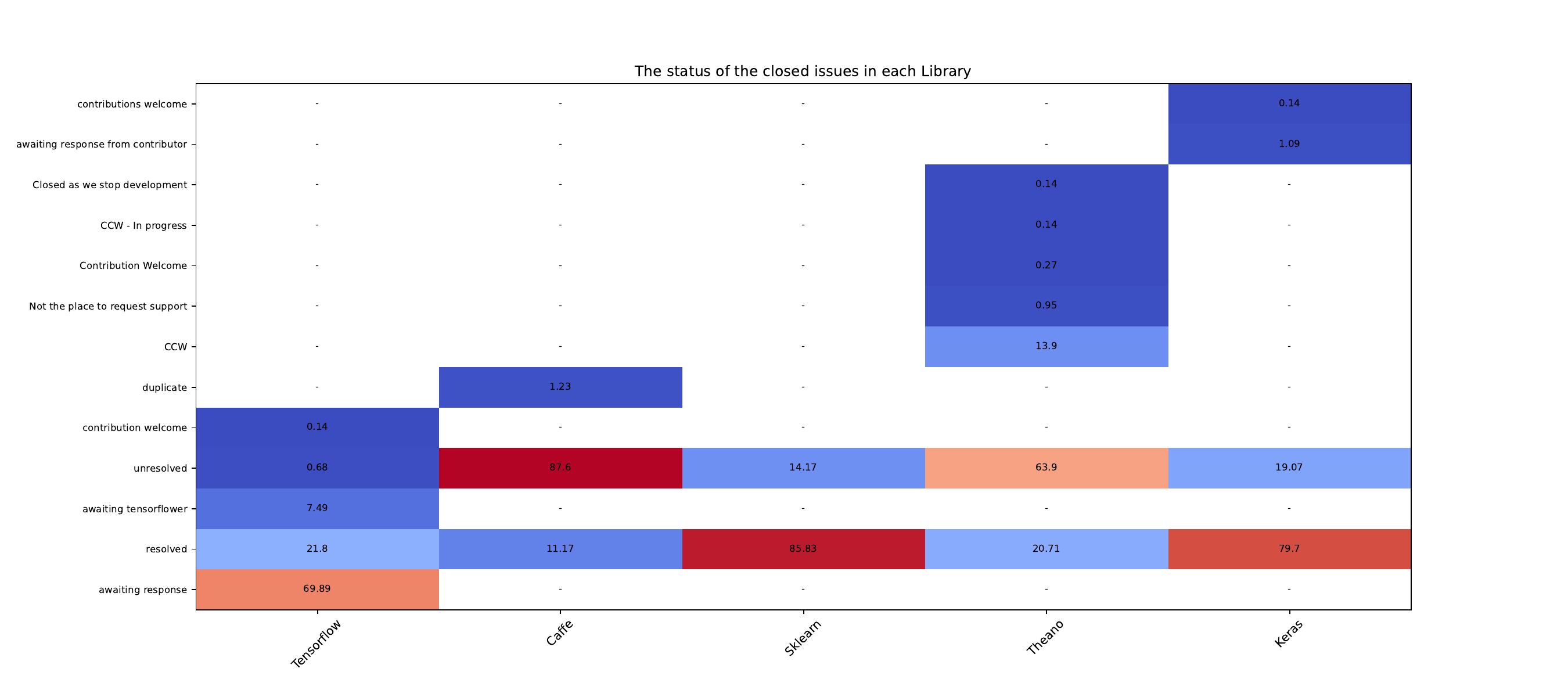}
\caption{Status of the closed issues in each library}
\label{RQ6visual1}
\end{figure*}

Similarly, \Cref{RQ6visual2} shows the status of the closed issues of five libraries: \te, \ca, \sk, \th, and \ke. It is evident that the majority of issues in \sk and \ke were resolved before closing, unlike in the other libraries, where only a few percentages of the closed issues were resolved. The high percentage of resolved issues suggests that development teams have the necessary technical expertise to identify and fix bugs or errors in library code. This may also indicate that they are actively engaging with the community and listening to user feedback, which can help improve the overall quality and usability of the libraries.

\begin{figure*}[ht]
\renewcommand\thefigure{VI}
\includegraphics[width=\textwidth,height=8cm]{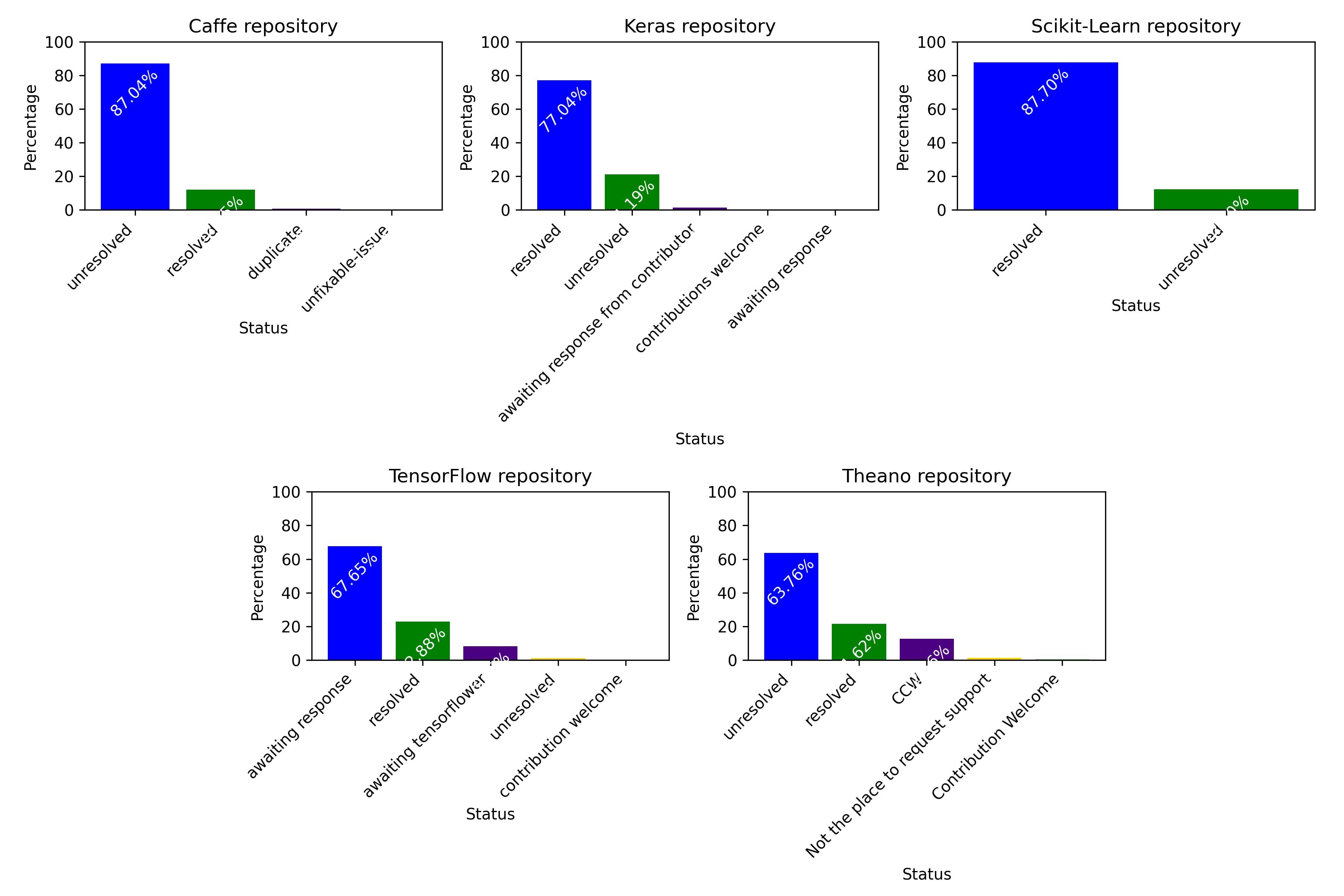}
\caption{Heatmap showing the relationship between status of issues in each library}
\label{RQ6visual2}
\end{figure*}

Furthermore, the majority of closed issues of \te \ml libraries were in the "awaiting response" status. However, RQ4 revealed that \te issues spend less time before receiving responses to the issues reported by the reporters. This suggests that there may be communication or workflow issues between the users who reported the issues and the development team responsible for addressing them.

The fact that the majority of closed issues of \te \ml libraries were in the "awaiting response" status does not necessarily mean that those issues were resolved or unresolved. It is possible that some of these issues were eventually resolved, but it is also possible that they remained unresolved due to a lack of communication or follow-up from either the user or development team.

Similarly, the majority of the reported issues on the \th and \ca repositories were not resolved. This suggests that there may be a need for additional resources, such as more developers or increased funding, to address the reported issues and improve library functionality and reliability. Alternatively, it may be necessary to re-prioritize the existing resources and focus on resolving the most critical issues first to ensure the users' satisfaction and confidence in the libraries. Understanding the reasons behind these unresolved issues can help identify potential solutions and improve the overall quality of libraries. This statement is supported by RQ1, which states that the majority of the reported issues in the two libraries were not categorized. Therefore, without proper categorization, it can be difficult for developers to identify the most pressing issues that need to be resolved. This can lead to a backlog of unresolved issues that can ultimately affect the usability and reliability of libraries. Therefore, it is important for developers to categorize the issues reported by users and prioritize their resolution based on their impact on library performance and functionality. Additionally, it is essential to allocate the necessary resources to address the reported issues and ensure that libraries meet users' needs and expectations.

To answer this RQ, the majority of the reported issues in some \ml libraries were resolved before closing, such as in \sk and \ke. However, others other issues were not resolved before closing, such as in \ca and \th, whereas some were waiting for responses, even though they were marked as closed, such as in \te.

\noindent \textbf{Implication:} The findings of this study suggest that different \ml libraries may have different issue-tracking processes and organizational levels. The effectiveness of the issue-tracking process can impact the speed and quality of issue resolution and, in turn, the overall user experience.

Libraries such as \sk and \ke, which have a high percentage of resolved issues before closing, may have a more streamlined and efficient issue-tracking process in place. Meanwhile, unresolved reported issues in \ca and \th may indicate the need for more formalized and organized issue-tracking procedures.

Additionally, the fact that some issues in \te were marked as closed but awaiting responses suggests that there may be room for improvement in communication and follow-up between users and the development team.

Overall, the issue tracking process is an important aspect of maintaining and improving \ml libraries, highlighting the need for libraries to have effective and well-organized issue tracking systems to ensure positive user experience.

\subsection{\RQSeven}
\noindent\textbf{Findings:} Correlation coefficient is a statistical measure that indicates the strength and direction of the relationship between two variables. In this case, the two variables being analyzed are the \textit{number of contributors} and the \textit{average duration of issue} for six different \ml libraries.

The correlation coefficient ranges from -1 to 1, where -1 indicates a strong negative correlation (i.e., when one variable increases, the other decreases), 0 indicates no correlation, and 1 indicates a strong positive correlation (i.e., when one variable increases, the other also increases).

\begin{figure*}[ht]
\renewcommand\thefigure{VII}
\includegraphics[width=\textwidth,height=8cm]{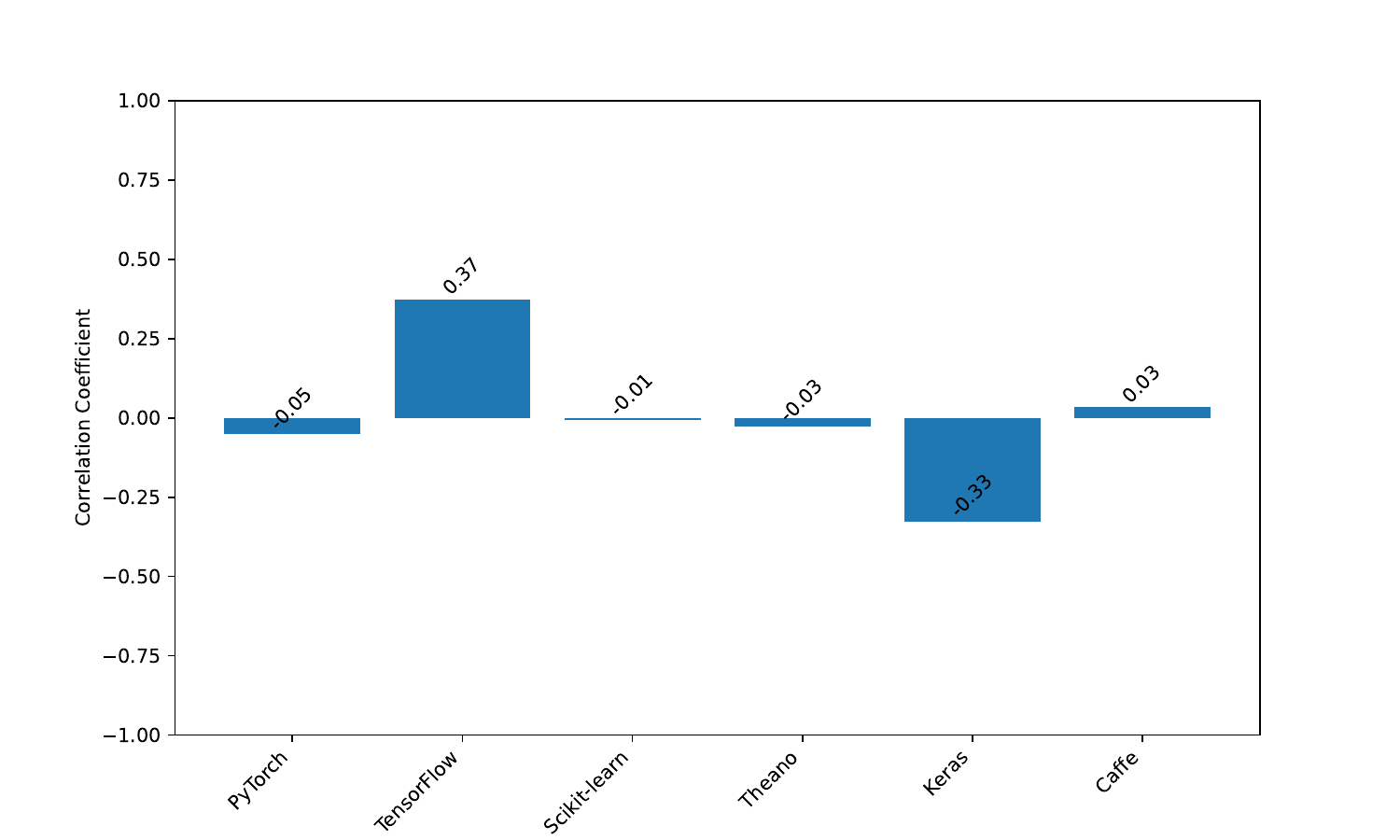}
\caption{Correlation coefficient between number of contributors and duration of issue}
\label{RQ7visual}
\end{figure*}

The results in \Cref{RQ7visual} showed that \te had a moderate positive correlation (0.37) between the number of contributors and the duration of issues, whereas \ke had a moderate negative correlation (-0.33). The moderate positive correlation between the number of contributors and the duration of issues in \te suggests that increasing the number of contributors may lead to longer issue-resolution times. This could be due to factors such as communication overheads or coordination difficulties. To address this issue, developers and maintainers of \te should consider implementing more effective communication channels, collaboration tools, and workflows to streamline the contribution process and reduce coordination costs. However, the moderate negative correlation between the number of contributors and the duration of issues in \ke suggests that increasing the number of contributors may lead to shorter issue resolution times. This could be due to factors such as increased diversity of skills and expertise, or the ability to leverage a larger pool of resources to solve issues. To take advantage of this effect, developers and maintainers of \ke could focus on building a more inclusive and diverse community as well as encouraging contributions from a wide range of skill levels and backgrounds

The remaining four libraries had very weak correlations (p $<$ 0.1) between these two variables. This suggests that there is little to no relationship between the number of contributors and the duration of issues in these libraries. The weak correlations between the number of contributors and the duration of issues suggest that the effect of the number of contributors on the issue resolution time is negligible. This could be due to various factors, such as the nature of the issues, development process, and size and complexity of the codebase. To improve the issue resolution times in these libraries, other factors such as documentation and testing may need to be improved.

To answer this research question, the analysis found that the number of contributors to an issue in \ml libraries has a varying impact on the time it takes to resolve the issue. While some libraries show a positive correlation, indicating longer issue resolution times with an increase in contributors, others show a negative correlation, indicating shorter issue resolution times with an increase in contributors. Effective communication and collaboration within the community are crucial to improving issue resolution times and the overall quality of \ml libraries.

\noindent\textbf{Implication:} The implications of these findings are that effective communication and collaboration within the community are crucial to improving issue resolution times and the overall quality of \ml libraries. Developers and maintainers of \ml libraries should consider implementing more effective communication channels, collaboration tools, and workflows to streamline the contribution process and reduce coordination costs. They should also focus on building a more inclusive and diverse community as well as encouraging contributions from a wide range of skill levels and backgrounds.

\section{Discussion}
\label{Sec::discussion}
In this session, we discuss the findings of our empirical study on the issues resolution and bug fixing process for six \ml libraries: \te, \ke, \th, \py, \ca, and \sk. Our study aimed to identify the most common software issues that arise in \ml libraries, as well as the most effective strategies for addressing these issues. We used quantitative methods to perform our analysis and answered seven research questions. The results of our study have significant implications for the development and maintenance of \ml libraries.
\subsection{Main Findings}
Our analysis revealed that "Bug," "Documentation," and "Performance" were the most common issues faced by users across all six \ml libraries. We found that developers should focus on improving the quality of their libraries by fixing critical bugs, optimizing performance, and improving documentation. Furthermore, developers can benefit from user engagement to gather feedback and prioritize bug fixes accordingly.

Our study also found that Testing and Runtime issues were the most commonly encountered challenges in all six libraries. These results can help developers prioritize development efforts and allocate resources more effectively to address the most common challenges faced by users.

Our OLS regression results showed that the \textit{total number of comments} on an issue has a significant impact on the time it takes to resolve the issue for all six \ml libraries. However, the R-squared values were low, indicating that the \textit{total number of comments} alone cannot explain much of the variation in the duration of the issues. The regression results suggest that libraries with larger coefficients for the \textit{total number of comments} tend to take longer to resolve issues when there are more comments.

We found that \te and \sk were highly responsive to user feedback, had a significantly lower average resolution time than the other libraries, and had well-defined categories of issues. Our study demonstrated the importance of responsiveness and efficient issue resolution in \ml libraries.

Our analysis revealed that it's essential to strike a balance between prioritizing critical issues and addressing other issues in a timely manner. Strategies to minimize the time spent on prioritized issues include thorough analysis and discussion before prioritizing, delegating issues to other contributors or community members, and improving communication and collaboration processes.

We also found that effective and well-organized issue tracking systems are crucial to ensuring a positive user experience. \sk and \ke had the highest percentage of resolved issues before closing, indicating efficient issue tracking processes. \te had the majority of issues in \textit{awaiting response} status, suggesting communication or workflow issues. \te and \ca had no labels to show the priority of reported issues, indicating less organized issue tracking processes.

Finally, our study revealed that effective communication and collaboration within a diverse community are crucial for improving issue resolution time and code quality in open-source software development.

\section{Threats to Validity}
\label{Sec::ttv}
\subsection{Construct Validity}
We conducted an empirical investigation of the issue resolution for effective bug-fixing process for six \ml libraries: \te, \ke, \th, \py, \ca, and \sk. Our goal was to identify the most common software issues that arise in \ml libraries and the most effective strategies for addressing these issues. However, there are certain threats to the construct validity of our study that have been mitigated.
Kalliamvakou \etal \citep{kalliamvakou2014promises} pointed out that many active projects do not conduct all of their software development activities on GitHub. Instead, separate infrastructure, such as mailing lists, forums, comments, etc., may be used to support decision-making processes. While this is a minor threat to our study, as most of the events of interest are recorded in the issue tracking system, it is important to note that some data may be missed if they are not recorded in the issue tracking system. Nonetheless, the issue tracking system is primarily used by development teams to track issue data, which means that the majority of the necessary data is likely to be recorded there. The data we focused on included the issue created date, issue closed date, issue duration, issue priority, issue status, issue title, issue topic, issue comment, issue contributors, and issue time of the first comment.
\subsection{Internal Validity}
The internal validity of this study pertains to the limitation of the analysis, which was based solely on keywords extracted from issue titles using the LDA algorithm. This approach may not capture all relevant information about the issues. However, we addressed this concern by validating the accuracy of the LDA algorithm by comparing its output with that of manually annotated data.
\subsection{Conclusion Validity}
The validity of our study could potentially be threatened by unconsidered confounding variables, such as the severity or complexity of the issues, since these variables are not available in the GitHub Issue repository. However, all necessary variables for this study were present in the dataset repository used in this study. Furthermore, we implemented various quantitative methods, including regression analysis, percentage and count analysis, and heatmap analysis, which fully captured the complexity of the data.
\subsection{External Validity}
A potential threat to the external validity of our study is sampling bias. Our findings may not be generalizable to other software domains or \ml libraries with different bug-fixing processes or user communities. However, we addressed this concern by using a large dataset with a variety of issue reports and keywords, which covered possible scenarios reported in other \ml libraries.

\section{Related Work}
\label{Sec::related}
First, we discuss the research on the phases involved in resolving issues, as the thrust of our investigation primarily revolves around the issue resolution processes. Specifically, we discuss the advancement of issue resolution and the interplay of certain metrics with one another, which leads to a more refined bug fixing process. (triaging, resolution, and integration). Subsequently, we discuss the extant research on the bug-fixing process.

\subsection{Issue Resolution Process}
Numerous studies have been conducted in software engineering to better understand the issue resolution process \citep{song2020empirical, murgia2014influence, bijlsma2012faster, lou2020understanding, kavaler2017perceived, nguyen2011impact, gonccalves2008implementing, yang2016recommending}. Before addressing an issue, the issue must be evaluated to determine if it's worth resolving and which developer should be responsible for resolving it. This process is known as "issue triaging" \citep{huang2019empirical}. A recent study has developed a multi-triage model to assign developers and allocate issue types in the bug triage process \citep{aung2022multi}. Additionally, Jiazhen et al. \citep{gu2020efficient} studied customer issue triaging in cloud services systems and discovered that linking customer incidents with detected system incidents help with customer incident triage.

Kim and Whitehead \citep{kim2006long} analyzed the time required to resolve issues by computing the bug-fix time of files in ArgoUML and PostgreSQL. They identified when bugs were introduced and when they were fixed. While some research has focused on estimating the time taken to resolve issue reports, such as Weiss et al. \citep{weiss2007long} and Zhang et al. \citep{zhang2013predicting}, which estimated the time taken to resolve issues based on the similarity between two reported issues, other studies have focused on the time required to fix bugs, such as Xuan et al. \citep{xuan2014towards} and Abdelmoez et al. \citep{abdelmoez2012bug}.

In the study by Guo et al. \citep{guo2010characterizing}, logistic regression was used to determine the likelihood that an issue report would be resolved. The experimental result showed a precision of 0.68 and a recall of 0.64. In some cases, a closed issue report may be re-opened if the solution is unsatisfactory for the issue reporter \citep{gu2010has, bhattacharya2013empirical, shihab2010predicting}. Zhongxian et al. \citep{gu2010has} found that bad fixes comprised as much as 9\% of all bugs in large bug databases of the Ant, AspectJ, and Rhino projects. Detecting and correcting bad fixes is critical for improving the quality and reliability of software. Similarly, Pamela et al. \citep{bhattacharya2013empirical} discovered that developer involvement decreased in some projects, even when bug-report quality was high, indicating that bug triaging is still problematic. Finally, Emad et al. \citep{shihab2010predicting} studied bugs that were initially closed but later reopened due to unsatisfactory resolutions or other reasons. They used the Eclipse project as a case study and found that comment and description text, the time it took to fix the bug, and the component the bug was found in were the most important factors in determining whether a bug would be reopened. They built explainable prediction models that could achieve 62.9\% precision and 84.5\% recall when predicting whether a bug would be reopened.

\subsection{Bug Fixing Process}
Bug fixing is a critical process in software development that involves resolving user-reported issues. The entire process is typically documented in bug reports, which track the life cycle of bugs from the time they are reported until they are resolved. Various studies have been conducted to improve the efficiency and effectiveness of the bug-fixing processes.

Sepahvand \etal \citep{sepahvand2022effective} proposed a model for predicting the number of code changes required to fix bugs. They approached this problem as a text classification task and used ensemble classifiers to develop their models. Statistical analysis revealed that there was a significant correlation between the number of code changes required and the time taken to fix the bugs. The average accuracy of the best model was approximately 72 

Similarly, Mariam \etal \citep{mezouar2018tweets} investigated the potential of social networks such as Twitter to improve the bug fixing process. They analyzed tweets posted by users of the Firefox and Chrome browsers to determine whether issue reports were treated differently based on whether they were mentioned on Twitter. Their experimental results showed that developers did not give special treatment to issues that were tweeted about unless the issue was labelled as more severe by the developers.

Finally, Renan \etal \citep{vieira2022bayesian} evaluated the impact of three report features: bug fix time (BFT), bug priority, code-churn size in bug fixing commits, and the presence of links to other reports on the time it takes to fix bugs. Using Bayesian statistics to analyze 55 projects from the Apache ecosystem, they found that higher code-churn values and the presence of links were associated with BFTs that were at least twice as long.

In summary, research has been conducted to improve the bug fixing and issue resolution processes in software development. Studies have focused on predicting the amount of code changes required, the potential of social networks in improving the process, and the impact of various report features on bug fix time, issue resolution time prediction, bug triaging type, and factors contributing to the re-opening of closed issues.

\section{Conclusion and Recommendation}
\label{Sec::conclusion}
\subsection{Conclusion}
In this study, we investigated the issue resolution process for effective bug fixing in six \ml libraries: \te, \ke, \th, PyTorch, \ca, and \sk. This study analyzed various metrics, including issue categories, average response time, issue duration, priority given to issues, number of comments on issues, number of contributors, and total issues generated from each \ml library. We used quantitative methods, such as regression, percentage and frequency counts, heatmaps, and correlation analysis, to analyze the reported issues in each library. The study results provide valuable insights into the most common issues that arise in \ml libraries, the effectiveness of their resolution, and the implications for the bug-fixing process.

The findings revealed that the most common issues faced by users of \ml libraries were related to bugs, documentation, and performance. The study also identified the most common issues across all six libraries, which were related to testing and runtime. Regression analysis showed that the total number of comments on an issue significantly affected its duration. Additionally, the study found that \te and \sk had the most responsive issue-tracking processes and the most efficient issue-resolution times.

Our study highlights the importance of effective communication and collaboration processes, efficient issue tracking processes, and the prioritization of critical issues in \ml libraries. Developers can benefit from user engagement to gather feedback and prioritize bug fixes accordingly. In summary, our study provides valuable insights into the issue resolution status of \ml libraries for effective bug fixing processes, which can help developers improve the quality of their libraries and provide better user experience. Further investigation is required to understand the relationship between the number of comments on an issue and the duration of the issue resolution. Although our study found a significant impact, the R-squared values were low, indicating that other factors may have contributed to the variation in duration. In addition, we plan to explore user engagement and feedback strategies in \ml libraries in more detail and identify strategies for improving issue resolution times. Finally, our study focused on \ml libraries written in Python, and we plan to investigate how the issue resolution processes for bug fixing differ in libraries written in other programming languages. Specifically, we aim to understand whether there are language-specific challenges that developers need to be aware of, and whether there are language-specific best practices for issue resolution and bug fixing.

\subsection{Recommendation}
This study has significant implications for the development and maintenance of \ml libraries. Therefore, the following recommendations are crucial to address the challenges highlighted in our analysis and to improve the overall quality and reliability of \ml libraries.

\subsubsection{For \ml Library Maintainers:}
\begin{itemize}
    \item The library maintainers should prioritize the fixing of critical bugs and improve the quality of documentation.
    \item Library maintainers should optimize the performance of the libraries to enhance user experience.
    \item They should engage users to gather feedback and prioritize bug fixes accordingly.
    \item They need to establish effective issue tracking systems to ensure efficient resolution of user-reported issues
\end{itemize}

\subsubsection{For \ml Library Users}
\begin{itemize}
    \item The users should provide clear and concise descriptions of the issues encountered to assist maintainers in fixing bugs and resolving issues
    \item The users should engage in open communication with the library maintainers to provide feedback and prioritize bug fixes
    \item The users should participate in testing and provide feedback to improve the performance of the libraries
\end{itemize}

\subsubsection{For Software Developers}
\begin{itemize}
    \item The developers should establish effective issue tracking systems to ensure efficient resolution of user-reported issues
    \item The developers should prioritize fixing critical bugs and optimizing the performance of libraries
    \item The developers should engage users to gather feedback and prioritize bug fixes accordingly
    \item The developers should encourage open communication between maintainers and users to provide feedback and resolve issues in a timely manner
\end{itemize}
These recommendations are based on the shortcomings identified by the study, and their implementation will improve the overall user experience and the efficiency of issue resolution in \ml libraries.\\

\section*{Acknowledgements} 
This research is supported by the Major Program of the National Natural Science Foundation of China (No.62192733, No.62192730).

\section*{\textbf{Statements and Declarations}} 
\subsection*{Author Contributions:} Conceptualization, A.A.; methodology, A.A.; software, A.A.; data curation, A.A., Y.D and H.Y; original draft preparation, A.A.; writing—review and editing, A.A.,Y.D and H.Y; visualization, A.A.; funding acquisition, Y.D. All authors have read and agreed to the published version of the manuscript.

\subsection*{\textbf{Competing Interests}} The authors declare no conflict of interest

\section*{\textbf{Data Availability}}
Supplementary materials, which have been archived on Bitbucket\footnote{https://bitbucket.org/eyinlojuoluwa/issue-resolution/downloads/}, include;
\begin{itemize}
    \item The generated dataset
    \item The scripts used to analyze the dataset
    \item The scripts used to generate the issues
\end{itemize}

\bibliographystyle{plainnat}
\bibliography{bibliography}

\begin{thebibliography}{59}
\providecommand{\natexlab}[1]{#1}
\providecommand{\url}[1]{\texttt{#1}}
\expandafter\ifx\csname urlstyle\endcsname\relax
  \providecommand{\doi}[1]{doi: #1}\else
  \providecommand{\doi}{doi: \begingroup \urlstyle{rm}\Url}\fi

\bibitem[Abd-Alrazaq et~al.(2020)Abd-Alrazaq, Alhuwail, Househ, Hamdi, Shah, et~al.]{abd2020top}
Alaa Abd-Alrazaq, Dari Alhuwail, Mowafa Househ, Mounir Hamdi, Zubair Shah, et~al.
\newblock Top concerns of tweeters during the covid-19 pandemic: infoveillance study.
\newblock \emph{Journal of medical Internet research}, 22\penalty0 (4):\penalty0 e19016, 2020.

\bibitem[Abdelmoez et~al.(2012)Abdelmoez, Kholief, and Elsalmy]{abdelmoez2012bug}
W~Abdelmoez, Mohamed Kholief, and Fayrouz~M Elsalmy.
\newblock Bug fix-time prediction model using na{\"\i}ve bayes classifier.
\newblock In \emph{2012 22nd International Conference on Computer Theory and Applications (ICCTA)}, pages 167--172. IEEE, 2012.

\bibitem[Aung et~al.(2022)Aung, Wan, Huo, and Sui]{aung2022multi}
Thazin Win~Win Aung, Yao Wan, Huan Huo, and Yulei Sui.
\newblock Multi-triage: A multi-task learning framework for bug triage.
\newblock \emph{Journal of Systems and Software}, 184:\penalty0 111133, 2022.

\bibitem[Bhattacharya et~al.(2013)Bhattacharya, Ulanova, Neamtiu, and Koduru]{bhattacharya2013empirical}
Pamela Bhattacharya, Liudmila Ulanova, Iulian Neamtiu, and Sai~Charan Koduru.
\newblock An empirical analysis of bug reports and bug fixing in open source android apps.
\newblock In \emph{2013 17th European Conference on Software Maintenance and Reengineering}, pages 133--143. IEEE, 2013.

\bibitem[Bijlsma et~al.(2012)Bijlsma, Ferreira, Luijten, and Visser]{bijlsma2012faster}
Dennis Bijlsma, Miguel~Alexandre Ferreira, Bart Luijten, and Joost Visser.
\newblock Faster issue resolution with higher technical quality of software.
\newblock \emph{Software quality journal}, 20:\penalty0 265--285, 2012.

\bibitem[Blei et~al.(2003)Blei, Ng, and Jordan]{Blei2003}
David~M. Blei, Andrew~Y. Ng, and Michael~I. Jordan.
\newblock Latent dirichlet allocation.
\newblock \emph{J. Mach. Learn. Res.}, 3\penalty0 (null):\penalty0 993–1022, mar 2003.
\newblock ISSN 1532-4435.

\bibitem[Brisson et~al.(2020)Brisson, Noei, and Lyons]{brisson2020we}
Scott Brisson, Ehsan Noei, and Kelly Lyons.
\newblock We are family: analyzing communication in github software repositories and their forks.
\newblock In \emph{2020 IEEE 27th International Conference on Software Analysis, Evolution and Reengineering (SANER)}, pages 59--69. IEEE, 2020.

\bibitem[Budhwani and Sun(2020)]{budhwani2020creating}
Henna Budhwani and Ruoyan Sun.
\newblock Creating covid-19 stigma by referencing the novel coronavirus as the “chinese virus” on twitter: quantitative analysis of social media data.
\newblock \emph{Journal of Medical Internet Research}, 22\penalty0 (5):\penalty0 e19301, 2020.

\bibitem[Bun and Harrison(2019)]{bun2019ols}
Maurice~JG Bun and Teresa~D Harrison.
\newblock Ols and iv estimation of regression models including endogenous interaction terms.
\newblock \emph{Econometric Reviews}, 38\penalty0 (7):\penalty0 814--827, 2019.

\bibitem[Bundschus et~al.(2009)Bundschus, Yu, Tresp, Rettinger, Dejori, and Kriegel]{bundschus2009hierarchical}
Markus Bundschus, Shipeng Yu, Volker Tresp, Achim Rettinger, Mathaeus Dejori, and Hans-Peter Kriegel.
\newblock Hierarchical bayesian models for collaborative tagging systems.
\newblock In \emph{2009 Ninth IEEE International Conference on Data Mining}, pages 728--733. IEEE, 2009.

\bibitem[Chen and Babar(2022)]{chen2022security}
Huaming Chen and M~Ali Babar.
\newblock Security for machine learning-based software systems: a survey of threats, practices and challenges.
\newblock \emph{arXiv preprint arXiv:2201.04736}, 2022.

\bibitem[Cheng et~al.(2018)Cheng, Cao, Xu, and Ma]{cheng2018manifesting}
Dawei Cheng, Chun Cao, Chang Xu, and Xiaoxing Ma.
\newblock Manifesting bugs in machine learning code: An explorative study with mutation testing.
\newblock In \emph{2018 IEEE International Conference on Software Quality, Reliability and Security (QRS)}, pages 313--324. IEEE, 2018.

\bibitem[Fr{\"a}drich et~al.(2020)Fr{\"a}drich, Oberm{\"u}ller, K{\"o}rber, Heuer, and Fraser]{fradrich2020common}
Christoph Fr{\"a}drich, Florian Oberm{\"u}ller, Nina K{\"o}rber, Ute Heuer, and Gordon Fraser.
\newblock Common bugs in scratch programs.
\newblock In \emph{Proceedings of the 2020 ACM conference on innovation and technology in computer science education}, pages 89--95, 2020.

\bibitem[Francalanci and Merlo(2008)]{francalanci2008empirical}
Chiara Francalanci and Francesco Merlo.
\newblock Empirical analysis of the bug fixing process in open source projects.
\newblock In \emph{Open Source Development, Communities and Quality: IFIP 20 th World Computer Congress, Working Group 2.3 on Open Source Software, September 7-10, 2008, Milano, Italy 4}, pages 187--196. Springer, 2008.

\bibitem[Gartziandia et~al.(2022)Gartziandia, Arrieta, Ayerdi, Illarramendi, Agirre, Sagardui, and Arratibel]{gartziandia2022machine}
Aitor Gartziandia, Aitor Arrieta, Jon Ayerdi, Miren Illarramendi, Aitor Agirre, Goiuria Sagardui, and Maite Arratibel.
\newblock Machine learning-based test oracles for performance testing of cyber-physical systems: An industrial case study on elevators dispatching algorithms.
\newblock \emph{Journal of Software: Evolution and Process}, 34\penalty0 (11):\penalty0 e2465, 2022.

\bibitem[Gheorghe et~al.(2020)Gheorghe, Gheorghe, and Iatan]{gheorghe2020agile}
Alina-M{\u{a}}d{\u{a}}lina Gheorghe, Ileana~Daniela Gheorghe, and Ioana~Laura Iatan.
\newblock Agile software development.
\newblock \emph{Informatica Economica}, 24\penalty0 (2), 2020.

\bibitem[Gon{\c{c}}alves et~al.(2008)Gon{\c{c}}alves, Bezerra, Belchior, Coelho, and Pires]{gonccalves2008implementing}
Fca M{\'a}rcia~GS Gon{\c{c}}alves, Carla Ilane~Moreira Bezerra, Arnaldo~Dias Belchior, Ciro~Carneiro Coelho, and Carlo Giovano~S Pires.
\newblock Implementing causal analysis and resolution in software development projects: The minidmaic approach.
\newblock In \emph{19th Australian Conference on Software Engineering (aswec 2008)}, pages 112--119. IEEE, 2008.

\bibitem[Goyal and Sardana(2019)]{goyal2019empirical}
Anjali Goyal and Neetu Sardana.
\newblock Empirical analysis of ensemble machine learning techniques for bug triaging.
\newblock In \emph{2019 Twelfth International Conference on Contemporary Computing (IC3)}, pages 1--6. IEEE, 2019.

\bibitem[Goyal and Sardana(2020)]{goyal2020performance}
Anjali Goyal and Neetu Sardana.
\newblock Performance assessment of bug fixing process in open source repositories.
\newblock \emph{Procedia Computer Science}, 167:\penalty0 2070--2079, 2020.

\bibitem[Gu et~al.(2020)Gu, Wen, Wang, Zhao, Luo, Kang, Zhou, Yang, Sun, Xu, et~al.]{gu2020efficient}
Jiazhen Gu, Jiaqi Wen, Zijian Wang, Pu~Zhao, Chuan Luo, Yu~Kang, Yangfan Zhou, Li~Yang, Jeffrey Sun, Zhangwei Xu, et~al.
\newblock Efficient customer incident triage via linking with system incidents.
\newblock In \emph{Proceedings of the 28th ACM Joint Meeting on European Software Engineering Conference and Symposium on the Foundations of Software Engineering}, pages 1296--1307, 2020.

\bibitem[Gu et~al.(2010)Gu, Barr, Hamilton, and Su]{gu2010has}
Zhongxian Gu, Earl~T Barr, David~J Hamilton, and Zhendong Su.
\newblock Has the bug really been fixed?
\newblock In \emph{Proceedings of the 32nd ACM/IEEE International Conference on Software Engineering-Volume 1}, pages 55--64, 2010.

\bibitem[Guo et~al.(2010)Guo, Zimmermann, Nagappan, and Murphy]{guo2010characterizing}
Philip~J Guo, Thomas Zimmermann, Nachiappan Nagappan, and Brendan Murphy.
\newblock Characterizing and predicting which bugs get fixed: an empirical study of microsoft windows.
\newblock In \emph{Proceedings of the 32Nd ACM/IEEE International Conference on Software Engineering-Volume 1}, pages 495--504, 2010.

\bibitem[Guo et~al.(2011)Guo, Zimmermann, Nagappan, and Murphy]{guo2011not}
Philip~J Guo, Thomas Zimmermann, Nachiappan Nagappan, and Brendan Murphy.
\newblock " not my bug!" and other reasons for software bug report reassignments.
\newblock In \emph{Proceedings of the ACM 2011 conference on Computer supported cooperative work}, pages 395--404, 2011.

\bibitem[Huang et~al.(2019)Huang, da~Costa, Zhang, and Zou]{huang2019empirical}
Yonghui Huang, Daniel~Alencar da~Costa, Feng Zhang, and Ying Zou.
\newblock An empirical study on the issue reports with questions raised during the issue resolving process.
\newblock \emph{Empirical Software Engineering}, 24:\penalty0 718--750, 2019.

\bibitem[Islam et~al.(2019)Islam, Nguyen, Pan, and Rajan]{islam2019comprehensive}
Md~Johirul Islam, Giang Nguyen, Rangeet Pan, and Hridesh Rajan.
\newblock A comprehensive study on deep learning bug characteristics.
\newblock In \emph{Proceedings of the 2019 27th ACM Joint Meeting on European Software Engineering Conference and Symposium on the Foundations of Software Engineering}, pages 510--520, 2019.

\bibitem[Kalliamvakou et~al.(2014)Kalliamvakou, Gousios, Blincoe, Singer, German, and Damian]{kalliamvakou2014promises}
Eirini Kalliamvakou, Georgios Gousios, Kelly Blincoe, Leif Singer, Daniel~M German, and Daniela Damian.
\newblock The promises and perils of mining github.
\newblock In \emph{Proceedings of the 11th working conference on mining software repositories}, pages 92--101, 2014.

\bibitem[Kaur and Kaur(2022)]{kaur2022empirical}
Harguneet Kaur and Arvinder Kaur.
\newblock An empirical study of aging related bug prediction using cross project in cloud oriented software.
\newblock \emph{Informatica}, 46\penalty0 (8), 2022.

\bibitem[Kavaler et~al.(2017)Kavaler, Sirovica, Hellendoorn, Aranovich, and Filkov]{kavaler2017perceived}
David Kavaler, Sasha Sirovica, Vincent Hellendoorn, Raul Aranovich, and Vladimir Filkov.
\newblock Perceived language complexity in github issue discussions and their effect on issue resolution.
\newblock In \emph{2017 32nd IEEE/ACM International Conference on Automated Software Engineering (ASE)}, pages 72--83. IEEE, 2017.

\bibitem[Kim and Whitehead~Jr(2006)]{kim2006long}
Sunghun Kim and E~James Whitehead~Jr.
\newblock How long did it take to fix bugs?
\newblock In \emph{Proceedings of the 2006 international workshop on Mining software repositories}, pages 173--174, 2006.

\bibitem[LaToza et~al.(2006)LaToza, Venolia, and DeLine]{latoza2006maintaining}
Thomas~D LaToza, Gina Venolia, and Robert DeLine.
\newblock Maintaining mental models: a study of developer work habits.
\newblock In \emph{Proceedings of the 28th international conference on Software engineering}, pages 492--501, 2006.

\bibitem[Li et~al.(2015)Li, Ouyang, Lu, Zhou, and Tian]{li2015group}
Ximing Li, Jihong Ouyang, You Lu, Xiaotang Zhou, and Tian Tian.
\newblock Group topic model: organizing topics into groups.
\newblock \emph{Information Retrieval Journal}, 18:\penalty0 1--25, 2015.

\bibitem[Liang et~al.(2014)Liang, Liu, Tan, and Bai]{liang2014sentiment}
Jiguang Liang, Ping Liu, Jianlong Tan, and Shuo Bai.
\newblock Sentiment classification based on as-lda model.
\newblock \emph{Procedia Computer Science}, 31:\penalty0 511--516, 2014.

\bibitem[Liao et~al.(2018)Liao, He, Chen, Fan, Zhang, and Liu]{liao2018exploring}
Zhifang Liao, Dayu He, Zhijie Chen, Xiaoping Fan, Yan Zhang, and Shengzong Liu.
\newblock Exploring the characteristics of issue-related behaviors in github using visualization techniques.
\newblock \emph{IEEE Access}, 6:\penalty0 24003--24015, 2018.

\bibitem[Liu et~al.(2013)Liu, Yang, Tan, and Hafiz]{liu2013r2fix}
Chen Liu, Jinqiu Yang, Lin Tan, and Munawar Hafiz.
\newblock R2fix: Automatically generating bug fixes from bug reports.
\newblock In \emph{2013 IEEE Sixth international conference on software testing, verification and validation}, pages 282--291. IEEE, 2013.

\bibitem[Lou et~al.(2020)Lou, Chen, Cao, Hao, and Zhang]{lou2020understanding}
Yiling Lou, Zhenpeng Chen, Yanbin Cao, Dan Hao, and Lu~Zhang.
\newblock Understanding build issue resolution in practice: symptoms and fix patterns.
\newblock In \emph{Proceedings of the 28th ACM Joint Meeting on European Software Engineering Conference and Symposium on the Foundations of Software Engineering}, pages 617--628, 2020.

\bibitem[Mezouar et~al.(2018)Mezouar, Zhang, and Zou]{mezouar2018tweets}
Mariam~El Mezouar, Feng Zhang, and Ying Zou.
\newblock Are tweets useful in the bug fixing process? an empirical study on firefox and chrome.
\newblock \emph{Empirical Software Engineering}, 23:\penalty0 1704--1742, 2018.

\bibitem[Murgia et~al.(2014)Murgia, Concas, Tonelli, Ortu, Demeyer, and Marchesi]{murgia2014influence}
Alessandro Murgia, Giulio Concas, Roberto Tonelli, Marco Ortu, Serge Demeyer, and Michele Marchesi.
\newblock On the influence of maintenance activity types on the issue resolution time.
\newblock In \emph{Proceedings of the 10th international conference on predictive models in software engineering}, pages 12--21, 2014.

\bibitem[Nguyen et~al.(2019)Nguyen, Dlugolinsky, Bob{\'a}k, Tran, L{\'o}pez~Garc{\'\i}a, Heredia, Mal{\'\i}k, and Hluch{\`y}]{nguyen2019machine}
Giang Nguyen, Stefan Dlugolinsky, Martin Bob{\'a}k, Viet Tran, {\'A}lvaro L{\'o}pez~Garc{\'\i}a, Ignacio Heredia, Peter Mal{\'\i}k, and Ladislav Hluch{\`y}.
\newblock Machine learning and deep learning frameworks and libraries for large-scale data mining: a survey.
\newblock \emph{Artificial Intelligence Review}, 52:\penalty0 77--124, 2019.

\bibitem[Nguyen et~al.(2010)Nguyen, Nguyen, Pham, Al-Kofahi, and Nguyen]{nguyen2010recurring}
Tung~Thanh Nguyen, Hoan~Anh Nguyen, Nam~H Pham, Jafar Al-Kofahi, and Tien~N Nguyen.
\newblock Recurring bug fixes in object-oriented programs.
\newblock In \emph{Proceedings of the 32nd ACM/IEEE International Conference on Software Engineering-Volume 1}, pages 315--324, 2010.

\bibitem[Nguyen~Duc et~al.(2011)Nguyen~Duc, Cruzes, Ayala, and Conradi]{nguyen2011impact}
Anh Nguyen~Duc, Daniela~S Cruzes, Claudia Ayala, and Reidar Conradi.
\newblock Impact of stakeholder type and collaboration on issue resolution time in oss projects.
\newblock In \emph{Open Source Systems: Grounding Research: 7th IFIP WG 2.13 International Conference, OSS 2011, Salvador, Brazil, October 6-7, 2011. Proceedings 7}, pages 1--16. Springer, 2011.

\bibitem[{\"O}zda{\u{g}}o{\u{g}}lu and Kavuncuba{\c{s}}{\i}(2019)]{ozdaugouglu2019monitoring}
G{\"u}zin {\"O}zda{\u{g}}o{\u{g}}lu and Ece Kavuncuba{\c{s}}{\i}.
\newblock Monitoring the software bug-fixing process through the process mining approach.
\newblock \emph{Journal of Software: Evolution and Process}, 31\penalty0 (7):\penalty0 e2162, 2019.

\bibitem[Pak et~al.(2020)Pak, Ma, Ryu, Ryom, Juhyok, Pak, and Pak]{pak2020deep}
Unjin Pak, Jun Ma, Unsok Ryu, Kwangchol Ryom, U~Juhyok, Kyongsok Pak, and Chanil Pak.
\newblock Deep learning-based pm2. 5 prediction considering the spatiotemporal correlations: A case study of beijing, china.
\newblock \emph{Science of The Total Environment}, 699:\penalty0 133561, 2020.

\bibitem[Panichella et~al.(2013)Panichella, Dit, Oliveto, Di~Penta, Poshynanyk, and De~Lucia]{panichella2013effectively}
Annibale Panichella, Bogdan Dit, Rocco Oliveto, Massimilano Di~Penta, Denys Poshynanyk, and Andrea De~Lucia.
\newblock How to effectively use topic models for software engineering tasks? an approach based on genetic algorithms.
\newblock In \emph{2013 35th International conference on software engineering (ICSE)}, pages 522--531. IEEE, 2013.

\bibitem[Saha et~al.(2015)Saha, Khurshid, and Perry]{saha2015understanding}
Ripon~K Saha, Sarfraz Khurshid, and Dewayne~E Perry.
\newblock Understanding the triaging and fixing processes of long lived bugs.
\newblock \emph{Information and software technology}, 65:\penalty0 114--128, 2015.

\bibitem[Schwarz(2018)]{schwarz2018ldagibbs}
Carlo Schwarz.
\newblock ldagibbs: A command for topic modeling in stata using latent dirichlet allocation.
\newblock \emph{The Stata Journal}, 18\penalty0 (1):\penalty0 101--117, 2018.

\bibitem[Sepahvand et~al.(2020)Sepahvand, Akbari, and Hashemi]{sepahvand2020predicting}
Reza Sepahvand, Reza Akbari, and Sattar Hashemi.
\newblock Predicting the bug fixing time using word embedding and deep long short term memories.
\newblock \emph{IET Software}, 14\penalty0 (3):\penalty0 203--212, 2020.

\bibitem[Sepahvand et~al.(2022)Sepahvand, Akbari, Hashemi, and Boushehrian]{sepahvand2022effective}
Reza Sepahvand, Reza Akbari, Sattar Hashemi, and Omid Boushehrian.
\newblock An effective model to predict the extension of code changes in bug fixing process using text classifiers.
\newblock \emph{Iranian Journal of Science and Technology, Transactions of Electrical Engineering}, pages 1--18, 2022.

\bibitem[Shihab et~al.(2010)Shihab, Ihara, Kamei, Ibrahim, Ohira, Adams, Hassan, and Matsumoto]{shihab2010predicting}
Emad Shihab, Akinori Ihara, Yasutaka Kamei, Walid~M Ibrahim, Masao Ohira, Bram Adams, Ahmed~E Hassan, and Ken-ichi Matsumoto.
\newblock Predicting re-opened bugs: A case study on the eclipse project.
\newblock In \emph{2010 17th Working Conference on Reverse Engineering}, pages 249--258. IEEE, 2010.

\bibitem[Song et~al.(2020)Song, Kong, Wang, and Li]{song2020empirical}
Qiwei Song, Xianglong Kong, Lulu Wang, and Bixin Li.
\newblock An empirical investigation into the effects of code comments on issue resolution.
\newblock In \emph{2020 IEEE 44th Annual Computers, Software, and Applications Conference (COMPSAC)}, pages 921--930. IEEE, 2020.

\bibitem[Thung et~al.(2012)Thung, Wang, Lo, and Jiang]{thung2012empirical}
Ferdian Thung, Shaowei Wang, David Lo, and Lingxiao Jiang.
\newblock An empirical study of bugs in machine learning systems.
\newblock In \emph{2012 IEEE 23rd International Symposium on Software Reliability Engineering}, pages 271--280. IEEE, 2012.

\bibitem[Vieira et~al.(2022)Vieira, Mesquita, Mattos, Britto, Rocha, and Gomes]{vieira2022bayesian}
Renan Vieira, Diego Mesquita, C{\'e}sar~Lincoln Mattos, Ricardo Britto, Lincoln Rocha, and Jo{\~a}o Gomes.
\newblock Bayesian analysis of bug-fixing time using report data.
\newblock In \emph{Proceedings of the 16th ACM/IEEE International Symposium on Empirical Software Engineering and Measurement}, pages 57--68, 2022.

\bibitem[Weiss et~al.(2007)Weiss, Premraj, Zimmermann, and Zeller]{weiss2007long}
Cathrin Weiss, Rahul Premraj, Thomas Zimmermann, and Andreas Zeller.
\newblock How long will it take to fix this bug?
\newblock In \emph{Fourth International Workshop on Mining Software Repositories (MSR'07: ICSE Workshops 2007)}, pages 1--1. IEEE, 2007.

\bibitem[Xuan et~al.(2014)Xuan, Jiang, Hu, Ren, Zou, Luo, and Wu]{xuan2014towards}
Jifeng Xuan, He~Jiang, Yan Hu, Zhilei Ren, Weiqin Zou, Zhongxuan Luo, and Xindong Wu.
\newblock Towards effective bug triage with software data reduction techniques.
\newblock \emph{IEEE transactions on knowledge and data engineering}, 27\penalty0 (1):\penalty0 264--280, 2014.

\bibitem[Yang et~al.(2016)Yang, Sun, Li, and Hu]{yang2016recommending}
Hui Yang, Xiaobing Sun, Bin Li, and Jiajun Hu.
\newblock Recommending developers with supplementary information for issue request resolution.
\newblock In \emph{Proceedings of the 38th International conference on software engineering companion}, pages 707--709, 2016.

\bibitem[Yu et~al.(2014)Yu, Mishra, and Mishra]{yu2014empirical}
Liguo Yu, Alok Mishra, and Deepti Mishra.
\newblock An empirical study of the dynamics of github repository and its impact on distributed software development.
\newblock In \emph{On the Move to Meaningful Internet Systems: OTM 2014 Workshops: Confederated International Workshops: OTM Academy, OTM Industry Case Studies Program, C\&TC, EI2N, INBAST, ISDE, META4eS, MSC and OnToContent 2014, Amantea, Italy, October 27-31, 2014. Proceedings}, pages 457--466. Springer, 2014.

\bibitem[Yu et~al.(2016)Yu, Wang, Yin, and Wang]{yu2016reviewer}
Yue Yu, Huaimin Wang, Gang Yin, and Tao Wang.
\newblock Reviewer recommendation for pull-requests in github: What can we learn from code review and bug assignment?
\newblock \emph{Information and Software Technology}, 74:\penalty0 204--218, 2016.

\bibitem[Zhang et~al.(2013)Zhang, Gong, and Versteeg]{zhang2013predicting}
Hongyu Zhang, Liang Gong, and Steve Versteeg.
\newblock Predicting bug-fixing time: an empirical study of commercial software projects.
\newblock In \emph{2013 35th International Conference on Software Engineering (ICSE)}, pages 1042--1051. IEEE, 2013.

\bibitem[Zhang et~al.(2015)Zhang, Wang, Hao, Xie, Zhang, and Mei]{zhang2015survey}
Jie Zhang, Xiaoyin Wang, Dan Hao, Bing Xie, Lu~Zhang, and Hong Mei.
\newblock A survey on bug-report analysis.
\newblock \emph{Sci. China Inf. Sci.}, 58\penalty0 (2):\penalty0 1--24, 2015.

\bibitem[Zhong and Su(2015)]{zhong2015empirical}
Hao Zhong and Zhendong Su.
\newblock An empirical study on real bug fixes.
\newblock In \emph{2015 IEEE/ACM 37th IEEE International Conference on Software Engineering}, volume~1, pages 913--923. IEEE, 2015.

\end{thebibliography}
\end{document}